\documentclass[12pt]{article}
\usepackage[utf8]{inputenc}
\usepackage{amssymb,amsmath,amsthm,mathtools}
\usepackage{dsfont}  %
\usepackage{graphicx}
\usepackage{enumerate}
\usepackage{natbib}
\usepackage{url}  %
\usepackage{xcolor}
\usepackage[shortcuts]{extdash}
\usepackage{caption,setspace}
\def\spacingset#1{\renewcommand{\baselinestretch}%
{#1}\small\normalsize} \spacingset{1}

\usepackage{xr}  %

\makeatletter
\newcommand*{\addFileDependency}[1]{%
\typeout{(#1)}%
\@addtofilelist{#1}
\IfFileExists{#1}{}{\typeout{No file #1.}}
}\makeatother

\newcommand*{\myexternaldocument}[1]{%
\externaldocument{#1}%
\addFileDependency{#1.tex}%
\addFileDependency{#1.aux}%
}

\myexternaldocument{suppMatb}

\newcommand{\Pb}{\mathbb{P}}
\newcommand{\Ev}{\mathbb{E}}
\newcommand{\Var}{\mathbb{V}}

\newcommand{\bB}{\boldsymbol{B}}
\newcommand{\bF}{\boldsymbol{F}}
\newcommand{\bP}{\boldsymbol{P}}
\newcommand{\by}{\boldsymbol{y}}
\newcommand{\bY}{\boldsymbol{Y}}
\newcommand{\bu}{\boldsymbol{u}}
\newcommand{\bU}{\boldsymbol{U}}
\newcommand{\bs}{\boldsymbol{s}}
\newcommand{\bI}{\boldsymbol{I}}
\newcommand{\bJ}{\boldsymbol{J}}

\newcommand{\btheta}{\boldsymbol{\theta}}

\newcommand{\diff}{\mathop{}\!\mathrm{d}}
\newcommand{\indFun}[1]{\mathds{1}_{\{#1\}}}

\newcommand{\blind}{1}

\title{\bf Nonparametric Two-Sample Test for Networks Using Joint Graphon Estimation}
\author{Benjamin Sischka\thanks{
    This work was partially funded by the BERD@NFDI Consortium (\texttt{https://www.berd-nfdi.de}).} ~and~G\"{o}ran Kauermann\\[0.1cm]
    Department of Statistics, Ludwig-Maximilians-Universit\"{a}t M\"{u}nchen
}

\begin{document}

\if1\blind
{
  \maketitle
} \fi

\bigskip
\begin{abstract}
    \noindent
    This paper focuses on the comparison of networks on the basis of statistical inference. For that purpose, we rely on smooth graphon models as a nonparametric modeling strategy that is able to capture complex structural patterns. The graphon itself can be viewed more broadly as density or intensity function on networks, making the model a natural choice for comparison purposes. Extending graphon estimation towards modeling multiple networks simultaneously consequently provides substantial information about the (dis-)sim\-i\-lar\-i\-ty between networks. Fitting such a joint model---which can be accomplished by applying an EM-type algorithm---provides a joint graphon estimate plus a corresponding prediction of the node positions for each network. In particular, it entails a generalized network alignment, where nearby nodes play similar structural roles in their respective domains. Given that, we construct a chi-squared test on equivalence of network structures. Simulation studies and real-world examples support the applicability of our network comparison strategy.
\end{abstract}

\noindent%
\vfill

\newpage
\spacingset{1.9} %

\section{Introduction}
The field of statistical modeling and analysis of complex networks has gained strongly increasing interest over the last two decades. This is driven by the fact that different types of systems can be reasonably formalized as relationships between individuals or interactions between objects. Analyzing such structures consequently allows to uncover and describe the phenomena that affect these systems. Network-structured data arise in various fields, for example, social and political sciences, economics, biology, neurosciences, and many others. In this regard, a connectivity pattern between entities might describe friendships among members of a social group \citep{eagle2009inferring}, the trading between nations \citep{bhattacharya2008international}, interactions of proteins \citep{schwikowski2000network}, or the functional coactivation within the human brain (\citealp{Bassett2018}, \citealp{Crossley2013}).

\begin{figure}%
	\centering
    \begin{minipage}[c]{0.49\textwidth}
        \centering
        \includegraphics[width=.99\textwidth]{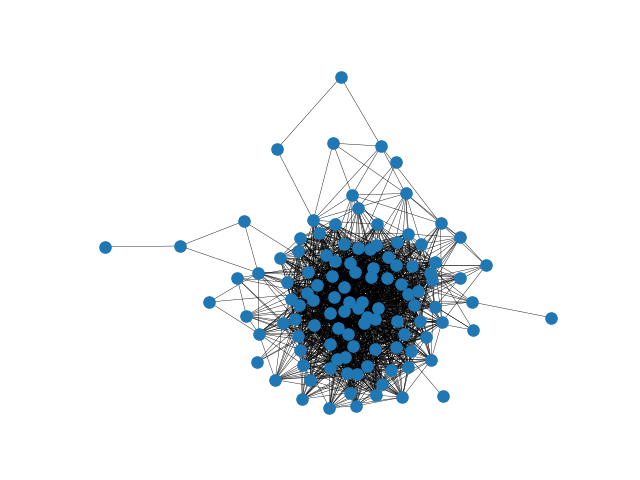}
    \end{minipage}
    \hfill
    \begin{minipage}[c]{0.49\textwidth}
        \centering
        \includegraphics[width=.99\textwidth]{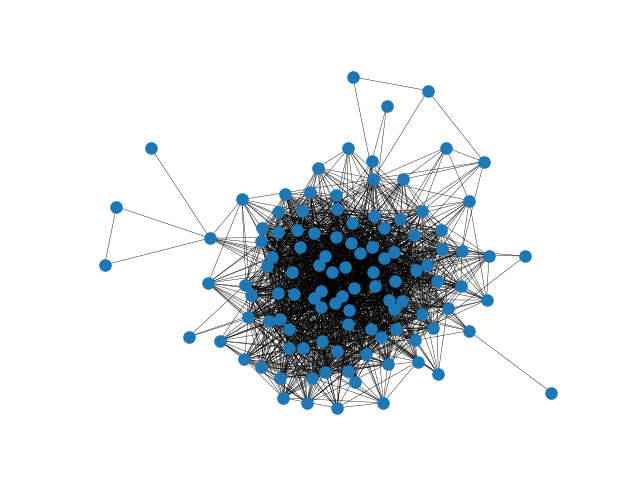}
    \end{minipage}
	\caption{Functional coactivation networks of the human brain. The illustrated connectivity patterns result from averaging over multiple measurements for subjects with autism spectrum disorder (left) and typical development (right). Do these networks reveal a significant structural difference?}
	\label{fig:brainIntro}
\end{figure}

In many situations, uncovering the underlying connectivity structure is not the only concern but also the comparison of akin networks and the exploration of potential differences. For example, this might be of interest in the context of brain coactivation. Recently, a lot of work has been going on investigating how the functional connectivity in the brain differs when people are affected by cognitive disorders like Alzheimer's disease or autism spectrum disorder (\citealp{song2019characterizing}, \citealp{subbaraju2017identifying}, \citealp{pascual2018evaluating}). Two such functional coactivation networks---resulting from respectively averaging over the measurements of two different subject groups---are illustrated in Figure~\ref{fig:brainIntro}. 
The posed research question in this context apparently also involves the inquiry of whether a significant difference in the brain processes is observable at all, which additionally might depend on the environmental conditions like resting state, external stimuli, etc. More generally, this can be phrased as a hypothesis test on structural equivalence of two networks.

To this end, we pursue constructing a model-based approach for network comparison that allows for formal statistical testing. More precisely, we aim to test whether two networks can be considered independent samples drawn from the same probability distribution. This is apparently in itself a technically difficult question since the two networks can have different sizes, and hence the two distributions need to be somehow different. Therefore, it is crucial that the applied distributional framework constitutes a rather universal probability measure. In fact, this is not a trivial property, and many network models entail conceptual issues that impede a direct comparison. For example, in the Exponential Random Graph Model \citep{robins2007introduction}, a concrete model parameterization has different implications for different network sizes, making corresponding coefficient estimates hardly comparable. Hence, it is necessary to rely on distributional models where the specification of size and edge probabilities can be explicitly disentangled.

For such a comparative analysis of networks, we will demonstrate that Graphon Models (\citealp{Lovasz2006}, \citealp{diaconis2007graph}) are a very useful tool. First, the graphon model is very flexible and able to capture complex network structures. Secondly, the graphon itself can be interpreted as nonparametric density or intensity function on networks. Both together make the graphon an overall characterizing network feature that, like nothing else, uniquely covers the structure in a comprehensive way. Hence, the graphon framework appears as a natural choice for comparison purposes. Lastly, the model's design fulfills the above requirement of decoupling the network's structure and size, allowing for modeling multiple networks simultaneously \citep{navarro2022joint}.

The rest of the paper is organized as follows. In Section~\ref{sec:netComLit}, we start with reviewing methods from the network comparison literature. A formalization of the test problem we want to tackle in this work is then concretely specified and discussed in Section~\ref{sec:notation}. The involved smooth graphon estimation in its capacity as the joint modeling approach is formulated in Section~\ref{sec:graphonEst}. Based on this joint graphon model, in Section~\ref{sec:testing}, we develop a network comparison strategy for testing on equivalence of the underlying structures. The general applicability of the complete approach is demonstrated in Section~\ref{sec:Appl}, where we consider its performance on simulated and real-world data. This involves the method's ability to uncover the underlying structure by joint graphon estimation, as well as the qualification of the subsequent testing procedure. The discussion and conclusion in Section~\ref{sec:discussion} completes the paper.

\section{Concepts for Network Comparison}  %
\label{sec:netComLit}
When reviewing the literature on network comparison, it is worth noting that all proposed strategies are naturally based on a specific concept of capturing network structures. In general, the various approaches available for comparing networks can be broadly distinguished according to whether they rely on a descriptive or a model-based structural framework. Survey articles in this field are given by \cite{Soundarajan2014}, \cite{Yaverolu2015}, \cite{Emmert-Streib2016}, and \cite{Tantardini2019}. A more general perspective on how complex data objects---such as adjacency matrices---might be compared is pointed out by \cite{marron2014overview}. Consulting this compendium clearly reveals a lack of model-based approaches in the context of network comparison. This is specifically deficient since drawing statistical inference is only possible under some kind of distributional assumption. The different strategies for network comparison proposed in the literature---irrespective of the capacity for drawing inference---are briefly reviewed hereafter.

Starting with approaches that are based on extracted network statistics, the most intuitive strategy is probably to simply compare global characteristics such as the clustering coefficient or the average path length (\citealp[pp.\ 364\, ff.]{newman2018networks}, \citealp[p.\ 31]{butts2008social}). However, this captures the overall network structure only very poorly since completely differently structured networks can apparently still possess the same global statistics. As a more advanced approach, \cite{Wilson2008} consider the differences in the graph spectra, see also \cite{Gera2018}. Yet, for the spectrum, it is often unclear which information in terms of local structural properties is extracted from the network. In addition, spectral methods can be strongly affected by small structural changes under specific circumstances. Taken together, such approaches often ascribe too little importance to the attributes of interest, leading to an over- or underrating of the structural dissimilarity at hand.

Another branch of the literature on descriptive network comparison relies on the concept of graphlets, i.e.\ prespecified subgraph patterns that are assumed to be sufficient for describing the present structure. Papers that, in one way or another, consider differences in the frequencies of graphlets are, among others, \cite{Przulj2004}, \cite{prvzulj2007biological}, \cite{ali2014alignment}, and \cite{faisal2017grafene}. Since the counting procedure is rather complex for larger graphlets, it is sort of a consensus to include only those that consist of no more than five nodes. However, this seems somehow arbitrary and incomplete in terms of capturing all structural aspects. Moreover, \cite{yaverouglu2014revealing} found high correlations among the graphlet-related statistics, including complete redundancies. In contrast, model-based approaches specify the complete distribution of frequencies over all kinds of subgraphs. A connection between subgraph frequencies and the concrete specification of the graphon model is exemplarily elaborated in \cite{borgs2008convergent}, \cite{bickel2011method}, and \cite{latouche2016variational}.

Overall, descriptive network statistics entail two general shortcomings. First, it is very difficult to assess in which parts of the networks the key differences are accommodated. To be precise, the nodes or edges (present or absent) that contribute most to a quantified structural discrepancy can only hardly be detected. Second, and more importantly, descriptive methods lack, by nature, the ability to draw inference in the probabilistic sense. Specifically, they provide no information on whether the found deviation between networks is plausible to be ascribed to randomness or whether there is a significant structural dissimilarity. \cite{butts2008social} aims to overcome this deficit by applying a simplistic conditional uniform graph distribution.

On the other hand, probabilistic models for network data allow to induce distributions on network patterns that extend to desired distributional assumptions on structural differences. As a consequence, these modeling approaches might potentially serve as structural construct used for comparison purposes. Yet, they possess individual conceptual shortcomings that often impede a direct comparison. While, for example, the Latent Distance Model \citep{Hoff2002} does not provide any model-related key component to be compared, coefficient estimates from the exponential random graph model are not directly comparable across separated networks. The graphon model and the Stochastic Blockmodel (\citealp{HollandLaskeyLein:83}, \citeauthor{Snijders1997}, \citeyear{Snijders1997} and \citeyear{Nowicki2001}) suffer from identifiability issues (see e.g.\ \citealp[Thm.\ 7.1]{diaconis2007graph}) that make a comparison of corresponding individual estimates complicated. The latter model's adaptivity is additionally highly dependent on the choice of the number of blocks. \cite{onnela2012taxonomies} tackle this issue by observing the networks' complete disintegration processes, which they subsequently summarize by the profiles of well-known network statistics. Integrating over the profiles' differences and applying principle component analysis for summarization yields the final distance measure, which has been demonstrated to provide reasonable results in terms of leading to a good classification. However, to the best of our knowledge, there exists no (model-based) \textit{nonparametric} test on the equivalence of network structures.

In this paper, we aim to address this shortcoming, which we tackle by striking new paths. As a general concept for this approach, we follow the intuition of fitting a joint model to multiple networks simultaneously. For that purpose, we resort to the smooth graphon model as an appropriate and very powerful framework. Such a joint modeling strategy consequently circumvents the need for post-hoc alignment of individual model fits and yields an outcome that provides substantial information for comparison purposes. More precisely, it allows for directly relating the networks at hand on the microscopic scale, which, in the literature, is often referred to as ``network alignment'' \citep{kuchaiev2010topological}. However, as an essential distinction to classical network alignment strategies, this method does not seek to find a node-wise one-to-one mapping. Instead, it implies a mapping of local components, meaning small fuzzy groups of nodes with similar structural roles in their respective domains. Based on this network alignment, a structural comparison at the microscopic level becomes possible. Aggregating local differences finally enables to construct a test on structural equivalence of networks.

\section{Notation and Formulation of the Test Problem}
\label{sec:notation}
We consider the setting where two undirected networks of possibly different sizes $N^{(1)}$ and $N^{(2)}$ have been observed. Let $\by^{(g)} = [y_{ij}^{(g)}]_{i,j = 1, \ldots, N^{(g)}}$ for $g =1, 2$ denote the two respective adjacency matrices, where, for $i,j =1, \ldots, N^{(g)}$, $y_{ij}^{(g)} = 1 $ if in network $g$ an edge between nodes $i$ and $j$ exists and $y^{(g)}_{ij}=0$ otherwise. That specifically means that $\by^{(g)} \in \{0,1\}^{N^{(g)} \times N^{(g)}}$. We assume the networks to be undirected so that $y_{ij}^{(g)} = y_{ji}^{(g)}$. Additionally, the diagonal elements are set to zero, i.e.\ $y_{ii}^{(g)}=0$, reflecting the absence of self-loops. In general, we consider $\by^{(g)}$ to be a realization of a random network $\bY^{(g)}$ of size $N^{(g)}$ which is subject to probability mass $\Pb(\bY^{(g)} = \by^{(g)} \, ; N^{(g)})$. The question we aim to tackle is whether $\by^{(1)}$ and $\by^{(2)}$ are drawn from the same distribution. To suitably specify such a distribution, we rely on the smooth graphon model. The data-generating process is thereby as follows. Assume that we independently draw uniformly distributed random variables
\begin{equation}
    U^{(g)}_i \sim \mbox{Uniform}[0,1] \quad \mbox{for } i = 1, \ldots, N^{(g)} \mbox{ and } g= 1,2.
    \label{eq:distOfU}
\end{equation}
Conditional on $\bU^{(g)}=(U^{(g)}_1, \ldots, U_{N^{(g)}}^{(g)})$, we then draw the edges i.i.d.\ through
\begin{equation}
    Y_{ij}^{(g)} \mid (\bU^{(g)} = \bu^{(g)} ) \sim \mbox{Binomial}(1 , w^{(g)}(u^{(g)}_i , u^{(g)}_j ) ) \quad \mbox{for } j>i
    \label{eq:dataGen}
\end{equation}
with $\bu^{(g)}=(u^{(g)}_1, \ldots, \allowbreak u_{N^{(g)}}^{(g)}) \in [0,1]^{N^{(g)}}$ and under the setting of $Y_{ij}^{(g)} \equiv Y_{ji}^{(g)}$ for $j<i$ and $Y_{ii}^{(g)} \equiv 0$. In this modeling framework, the function $w^{(g)}: [0,1]^2 \rightarrow [0,1]$, which specifies the structural behavior of the emerging network, is called graphon (see \citealp{Lovasz2006} and \citealp{diaconis2007graph}). Here, in particular, we assume $w^{(g)}(\cdot,\cdot)$ to be smooth according to some H\"older or Lipschitz condition (cf.\ \citealp{wolfe2013nonparametric} or \citealp{Chan2014}). Relying on this data-generating process, the graphon-based probability model can be defined through
\begin{equation}
    \bY^{(g)} \sim \Pb\big(\bY^{(g)} = \cdot \, ; \, w^{(g)}(\cdot, \cdot) , N^{(g)}\big).
    \label{eq:distMod}
\end{equation}
In this distribution model, the network's size and structure are apparently dissociated, which therefore allows for a size-independent comparison of underlying structures. Hence, our goal is to develop a statistical test on the hypothesis
\begin{equation}
    H_0 : \; w^{(1)}(\cdot, \cdot ) \equiv w^{(2)}(\cdot, \cdot).
    \label{eq:h0}
\end{equation}
In this context, we emphasize that data-generating process~(\ref{eq:dataGen}) is not unique because it is invariant to permutations of $w^{(g)}(\cdot,\cdot)$, as discussed in detail by \citet[Sec.\ 7]{diaconis2007graph}. Thus, the formulation of $H_0$ needs to be understood from the perspective of corresponding equivalence classes, implying that $w^{(1)}(\cdot, \cdot )$ and $w^{(2)}(\cdot, \cdot)$ are rather viewed from a theoretical perspective. Nonetheless, for the concrete implementation of the test procedure, we employ a concrete representation of the two graphons. Specifically, under the assumption of $H_0$ being true, we call the coinciding manifestation the \textit{joint graphon}. This can be formalized as
\begin{equation*}
    w^{\text{joint}}(u,v) := w^{(1)}(u, v) = w^{(2)}(u, v) \quad \mbox{for all } (u,v)^\top \in [0,1]^2.
\end{equation*}
Since $w^{(1)}(\cdot,\cdot)$ and $w^{(2)}(\cdot,\cdot)$ are assumed to be smooth---at least for one possible arrangement, and, in particular, the one we consider here---, this also holds for $w^{\text{joint}}(\cdot,\cdot)$. Given such a concrete representation of the joint graphon, the node position vectors $\bu^{(1)}$ and $\bu^{(2)}$ referring to $w^{\text{joint}}(\cdot,\cdot)$ then provide a specific type of network alignment. This is what we utilize for a direct comparison of $\by^{(1)}$ and $\by^{(2)}$. However, one typically observes neither $\bu^{(1)}$ and $\bu^{(2)}$ nor $w^{\text{joint}}(\cdot,\cdot)$. Thus, in order to achieve this alignment, we first need to formulate an appropriate estimation procedure for the joint graphon model.

\section{EM-Based Joint Graphon Estimation}
\label{sec:graphonEst}
In this section, we present an iterative estimation procedure for the joint smooth graphon model under the assumption that null hypothesis~(\ref{eq:h0}) is true. To do so, we follow the EM-based estimation routine of \cite{sischka2022based}, extending it to the situation of two networks.

\subsection{MCMC E-Step}
\label{sec:e-step}
Starting with the E-step of our iterative algorithm, we assume the joint graphon $w^{\text{joint}}(\cdot , \cdot)$ to be known for the moment. Based on that, the latent positions of the networks can be separately determined using MCMC techniques. To be precise, we apply Gibbs sampling by formulating the full conditional distribution of $U_i^{(g)}$ through
\begin{multline}
    f(u_i^{(g)} \mid u_1^{(g)},\ldots, u_{i-1}^{(g)},u_{i+1}^{(g)}, \ldots, u_{N^{(g)}}^{(g)}, \by^{(g)}) \\
    \propto \prod_{j \neq i} w^{\text{joint}}(u_i^{(g)} , u_j^{(g)})^{y_{ij}^{(g)}} [1 - w^{\text{joint}}(u_i^{(g)} , u_j^{(g)})]^{1-{y_{ij}^{(g)}}}
    \label{eq:gibbs}
\end{multline}
for all $i = 1,\ldots, N^{(g)}$ and $g=1, 2$. Details on the concrete implementation of the Gibbs sampler are given in Section~\ref{sec:GibbsApp} of the Appendix. The resulting MCMC sequence (after cutting the burn-in period and appropriate thinning) then reflects the joint conditional distribution $f(\bu^{(g)} \mid \by^{(g)})$. Thus, the marginal conditional means of the latent positions, i.e.\  $\Ev (U_i^{(g)} \mid \bY^{(g)} = \by^{(g)})$ for $i=1,\ldots,N^{(g)}$, can be approximated by taking the mean over the MCMC samples, which we denote by $\bar{\bu}^{(g)} = (\bar{u}_1^{(g)}, \ldots, \bar{u}_{N^{(g)}}^{(g)})$. This posterior mean vector, however, requires further adjustments due to additional identifiability issues which cannot be coped with the standard EM-type algorithm. To illustrate this, let model assumption~(\ref{eq:distOfU}) be more relaxed in the sense that the $U_i^{(g)}$'s might follow any continuous distribution $F^{(g)}(\cdot)$. Under this configuration, the model ($F^{(g)}(\cdot)$, $w^{(g)}(\cdot,\cdot)$) is equivalent to any other model (${F^{(g)}}^\prime(\cdot)$, ${w^{(g)}}^\prime(\cdot,\cdot)$) constructed through
\begin{equation*}
    {F^{(g)}}^\prime(u^\prime) := F^{(g)}(\varphi(u^\prime)) \quad \mbox{and} \quad {w^{(g)}}^\prime(u^\prime,v^\prime) := w^{(g)} (\varphi(u^\prime),\varphi(v^\prime))
\end{equation*}
with $\varphi: [0,1] \rightarrow [0,1]$ being a strictly increasing continuous function (that is, in contrast to \citealp[Sec.\ 7]{diaconis2007graph}, not measure-preserving). Specifically, that means
\[
    \Pb\big(\bY^{(g)} = \cdot \, ; \, F^{(g)}(\cdot), \allowbreak w^{(g)}(\cdot,\cdot) , \allowbreak N^{(g)}\big) \equiv \Pb\big(\bY^{(g)} = \cdot \, ; \, {F^{(g)}}^\prime(\cdot), \allowbreak {w^{(g)}}^\prime(\cdot,\cdot) , \allowbreak N^{(g)}\big) 
\]
for all $N^{(g)} \geq 2$. As a matter of conception, this issue cannot be solved by the EM algorithm since it aims at specifying a model that adapts optimally to the given data instead of perfectly recovering the underlying model structure. Consequently, the EM approach is not able to distinguish between the two conceptually equivalent model specifications ($F^{(g)}(\cdot)$, $w^{(g)}(\cdot,\cdot)$) and (${F^{(g)}}^\prime(\cdot)$, ${w^{(g)}}^\prime(\cdot,\cdot)$). Nonetheless, this identifiability issue can simply be tackled by adjusting $\bar{\bu}^{(g)}$ before estimating the graphon in the M-step. To do so, we just impose that the inferred node positions follow an ideal sample drawn from the standard uniform distribution. That is, we set
\begin{equation*}
    \hat{u}_i^{(g)} = \frac{\operatorname{rank} (\bar{u}_i^{(g)})}{N^{(g)}+1},
\end{equation*}
where $\operatorname{rank}(\bar{u}_i^{(g)})$ is the rank from smallest to largest of element $\bar{u}_i^{(g)}$ within $\bar{\bu}^{(g)}$. In this context, note that the values $i/(N^{(g)}+1)$ with $i=1,\ldots,N^{(g)}$ represent the expectations of $N^{(g)}$ ordered random variables that are independently drawn from the standard uniform distribution. As a result, with $\hat{\bu}^{(g)} = (\hat{u}_1^{(g)}, \ldots, \hat{u}_{N^{(g)}}^{(g)})$ we obtain a plausible realization of the node positions of network $g$. Apparently, this relies on the current joint graphon estimate $\hat{w}^{\text{joint}}(\cdot, \cdot)$, which is applied as substitute in conditional distribution~(\ref{eq:gibbs}). In the next step, we formulate the procedure for updating $\hat{w}^{\text{joint}}(\cdot, \cdot)$ given $\hat{\bu}^{(1)}$ and $\hat{\bu}^{(2)}$.

\subsection{Spline-Based M-Step}
\label{sec:spline}
For a semiparametric estimation of the joint smooth graphon, we choose a linear B-spline regression approach. To this end, we assume the joint graphon to be approximated through
\begin{equation*}
    w_{\btheta}^{\text{joint}}(u,v) = \bB(u,v) \, \btheta = \left[ \bB(u) \otimes \bB(v) \right] \btheta,
\end{equation*}
where $\otimes$ is the Kronecker product, $\bB(u) \in \mathbb{R}^{1 \times L}$ is a linear B-spline basis on $[0,1]$, normalized to have a maximum value of one, and $\btheta \in \mathbb{R}^{L^2}$ is the parameter vector to be estimated. The inner B-spline knots are specified as lying equidistantly on a regular 2D grid within $[0,1]^2$, where $\btheta$ is indexed accordingly through $\btheta = \left( \theta_{11},\ldots, \theta_{1L}, \allowbreak \theta_{21}, \ldots, \theta_{LL} \right)^\top$. Based on this representation and given the node positions $\hat{\bu}^{(1)}$ and $\hat{\bu}^{(2)}$, we formulate the marginal log-likelihood over both networks as
\begin{equation}
    \ell(\btheta) = \sum_g \sum\limits_{\substack{i,j \\ j \neq i}} \left[ y_{ij}^{(g)} \, \log \left( \bB_{ij}^{(g)} \btheta \right) + \left( 1- y_{ij}^{(g)} \right) \, \log \left( 1 - \bB_{ij}^{(g)} \btheta \right) \right],
    \label{eq:logLik}
\end{equation}
where $\bB_{ij}^{(g)} = \bB(\hat{u}_i^{(g)}) \otimes \bB(\hat{u}_j^{(g)})$. Furthermore, through standard calculations, we are able to derive the score function $\bs(\btheta)$ and the Fisher information $\bF(\btheta)$, as demonstrated in Section~\ref{sec:splineApp} of the Appendix. Fisher scoring can then be used to maximize $\ell(\btheta)$. In addition, we include side constraints to ensure that $w_{\btheta}^{\text{joint}}(\cdot,\cdot)$ is bounded to $[0,1]$ and symmetric. In the linear B-spline setting, this means restricting the parameters by the conditions
\begin{equation*}
    \theta_{kl} \geq 0 \; , \quad \theta_{kl} \leq 1 \; , \quad \mbox{and} \quad \theta_{kl} - \theta_{lk} = 0
\end{equation*}
for all $l > k$. Apparently, all three conditions are of linear form and thus can be written in matrix format. Taken together, the Fisher scoring becomes a quadratic programming problem that can be solved using standard software (see \citealp{cvxopt} or \citealp{quadprog}).

Moreover, we intend to add penalization on the B-spline estimate. As outlined in \cite{Eilers1996} and \citeauthor{Ruppert2003} (\citeyear{Ruppert2003}, \citeyear{Ruppert2009}), penalized spline estimation under the setting of a rather large spline basis yields a preferable outcome since it guarantees a functional fit that covers the data adequately but is still smooth. Thus, this approach enables to precisely capture the underlying structure while avoiding overfitting. To realize this, we add a first-order penalty, meaning that ``neighboring'' elements of $\btheta$ get penalized. For the log-likelihood, the score function, and the Fisher information, this leads to the penalized versions in the form of
\begin{equation}
    \begin{gathered}
        \ell_{\text{p}} (\btheta, \lambda) = \ell(\btheta) - \frac{1}{2} \lambda \btheta^\top \bP \btheta\, , \quad \bs_{\text{p}}(\btheta, \lambda) = \bs(\btheta) - \lambda \bP \btheta\, , \\ 
        \text{and} \quad \bF_{\text{p}}(\btheta, \lambda) = \bF (\btheta) + \lambda \bP,
    \end{gathered}
    \label{eq:penFuncs}
\end{equation}
respectively, where $\bP$ is a penalization matrix of appropriate shape (see Section~\ref{sec:splineApp} of the Appendix). For an adequate choice of the penalty parameter $\lambda$ in the two\-/dimensional spline regression, we follow \cite{Kauermann2013} and apply the corrected Akaike Information Criterion ($\operatorname{AIC}_{\text{c}}$, see \citealp{Hurvich1989} and \citealp{Burnham2002}). This is defined as
\begin{equation*}
    \operatorname{AIC}_{\text{c}} (\lambda) = -2 \, \ell(\hat{\btheta}_{\text{p}}) + 2 \, \operatorname{df} (\lambda) + \frac{2 \, \operatorname{df} (\lambda) [ \operatorname{df} (\lambda) +1 ] }{N(N-1) - \operatorname{df} (\lambda) -1},
\end{equation*}
where $\hat{\btheta}_{\text{p}}$ is the corresponding penalized parameter estimate and $\operatorname{df} (\lambda)$ specifies the degrees of freedom of the penalized B-spline function. More precisely, according to \citet[pp.\ 211\, ff.]{wood2017generalized}, the latter is defined trough
\begin{equation*}
    \operatorname{df} (\lambda) = \operatorname{tr} \left\{ \bF_{\text{p}}^{-1} (\hat{\btheta}_{\text{p}}, \lambda) \, \bF(\hat{\btheta}_{\text{p}}) \right\}
\end{equation*}
with $\operatorname{tr} \{\cdot \}$ being the trace of a matrix. A numerical optimization of the corrected $\operatorname{AIC}$ with respect to $\lambda$ concludes the estimation of $\btheta$, resulting in the eventual estimate $\hat{w}^{\text{joint}}(\cdot, \cdot)$ of the current M-step.

Finally, the EM-type estimation procedure described above---meaning the consecutive repetition of the E- and M-step until convergence is achieved---allows us to adequately estimate both the joint graphon $w^{\text{joint}}(\cdot,\cdot)$ and the corresponding node positions $\bu^{(1)}$ and $\bu^{(2)}$ of the two networks. Based on these results, we are now able to formulate an appropriate test procedure.

\section{Two-Sample Test on Network Structures}
\label{sec:testing}
\begin{figure}
    \centering
    \begin{minipage}[t]{0.32\textwidth}
        \centering
        \includegraphics[trim={45 0 60 0},clip,width=1\textwidth]{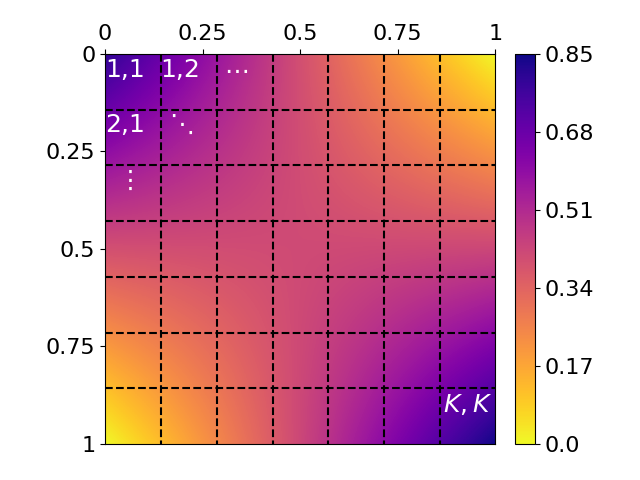}
    \end{minipage}
    \hfill
    \begin{minipage}[t]{0.32\textwidth}
        \centering
        \includegraphics[trim={45 0 60 0},clip,width=1\textwidth]{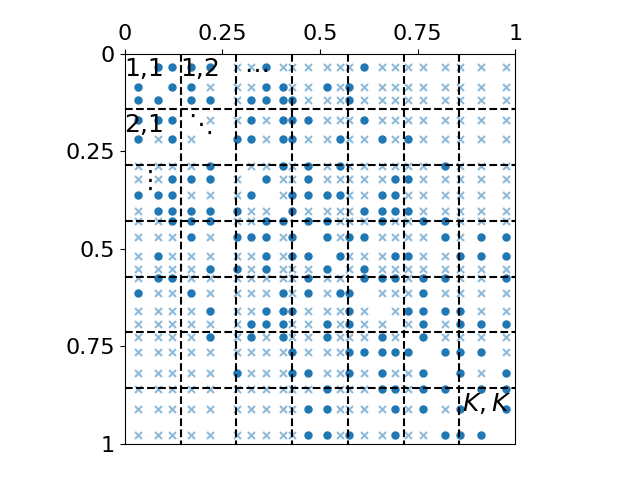}
    \end{minipage}
    \hspace*{-0.2cm}
    \begin{minipage}[t]{0.32\textwidth}
        \centering
        \includegraphics[trim={45 0 60 0},clip,width=1\textwidth]{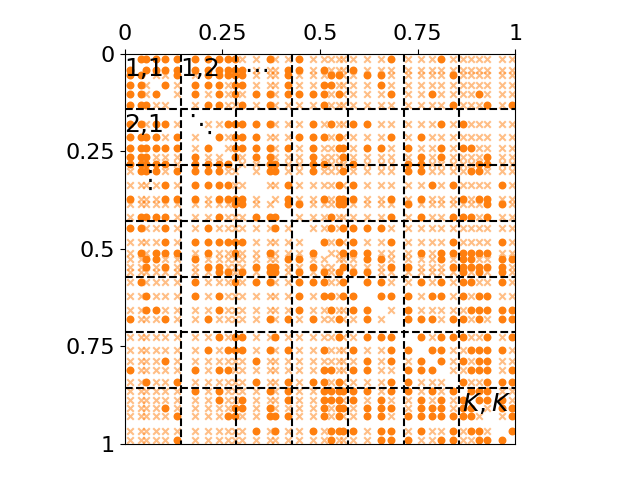}
    \end{minipage}
    \caption{Dividing the unit square as domain of the graphon model into small segments for comparing network structure on a microscopic level. Left: division of $w^{\text{joint}}(\cdot,\cdot)$ into approximately piecewise-constant rectangles. Middle and right: edge positions $(u_{i}^{(g)}, u_{j}^{(g)})^\top$ of two simulated networks with respect to $w^{\text{joint}}(\cdot,\cdot)$; weakly colored crosses and intensively colored circles represent absent and present edges, respectively. The two networks can be compared by pairwise contrasting the edge proportions within the labeled rectangles.}
    \label{fig:divideModel}
\end{figure}

Returning to the test problem raised in Section~\ref{sec:notation}, we now develop a statistical test procedure on hypothesis \eqref{eq:h0}, i.e.\ whether $\by^{(1)}$ and $\by^{(2)}$ are drawn from the same distribution. To do so, we utilize the network alignment resulting from the (inferred) joint smooth graphon model. More precisely, we exploit the fact that two edge variables $Y_{i_1 j_1}^{(1)}$ and $Y_{i_2 j_2}^{(2)}$ that have nearby positions---i.e.\ for which the distance between $(u_{i_1}^{(1)}, u_{j_1}^{(1)})^\top$ and $(u_{i_2}^{(2)}, u_{j_2}^{(2)})^\top$ is small---possess similar probabilities to form a connection. In a more formalized way, this means that, from $\Vert (u_{i_1}^{(1)}, u_{j_1}^{(1)})^\top - (u_{i_2}^{(2)}, u_{j_2}^{(2)})^\top \Vert \approx 0$, it follows that
\begin{multline*}
    \Pb(Y_{i_1 j_1}^{(1)} = 1 \mid U_{i_1}^{(1)} = u_{i_1}^{(1)}, U_{j_1}^{(1)} = u_{j_1}^{(1)} ) \\
    \approx \Pb(Y_{i_2 j_2}^{(2)} = 1 \mid U_{i_2}^{(2)} = u_{i_2}^{(2)}, U_{j_2}^{(2)} = u_{j_2}^{(2)} ), 
\end{multline*}
where $\Vert \cdot \Vert$ is the Euclidean distance. Following this intuition, we divide the unit square into small segments and compare between networks the ratio of present versus absent edges occurring in these segments (see Figure~\ref{fig:divideModel} for an exemplary division). 
For that purpose, we choose a suitable $K \in \mathbb{N}$, specify a corresponding boundary sequence $a_0=0 < a_1 < \ldots < a_K=1$, and define the following two quantities for $l,k = 1,\ldots,K$ with $l \geq k$:
\begin{align}
\begin{split}
    d_{kl}^{(g)} &= \sum\limits_{\substack{i,j \\ j > i}} y_{ij}^{(g)} \indFun{u_i^{(g)} \in [a_{k-1}, a_k)} \indFun{u_j^{(g)} \in [a_{l-1}, a_l)} \\
    m_{kl}^{(g)} &= \sum\limits_{\substack{i,j \\ j > i}} \indFun{u_i^{(g)} \in [a_{k-1}, a_k)} \indFun{u_j^{(g)} \in [a_{l-1}, a_l)}.
\end{split}
\label{eq:testQuants}
\end{align}
This means $d_{kl}^{(g)}$ and $m_{kl}^{(g)}$ represent the number of present ($y_{ij}^{(g)} = 1$) and a priori potential ($y_{ij}^{(g)} \in \{0,1\}$) edges of network $g$, respectively, within the constructed rectangle $[a_{k-1}, a_k) \times [a_{l-1}, a_l)$. The corresponding cross-network counts can be calculated by $d_{kl} = d_{kl}^{(1)} + d_{kl}^{(2)}$ and $m_{kl} = m_{kl}^{(1)} + m_{kl}^{(2)}$. Since $w^{\text{joint}}(\cdot,\cdot)$ is smooth, we further assume that the induced probability on edge variables within $[a_{k-1}, a_k) \times [a_{l-1}, a_l)$ is approximately constant. That allows for putting the observed ratios between present and absent edges in direct relation. In this light, we formulate the following contingency table to keep track of homogeneity between the networks within rectangle $(k,l)$:
\begin{center}
\begin{tabular}{|cc||c|}
    \hline
    $d_{kl}^{(1)}$ & $d_{kl}^{(2)}$ & $d_{kl}$ \\
    $m_{kl}^{(1)} - d_{kl}^{(1)}$ & $m_{kl}^{(2)} - d_{kl}^{(2)}$ & $m_{kl} - d_{kl}$  \\ \hline \hline
    $m_{kl}^{(1)}$ & $m_{kl}^{(2)}$ & $m_{kl}$ \\
    \hline
\end{tabular}
\end{center}
Apparently, if $H_0$ is assumed to be true, we would expect the proportions of present edges, $d_{kl}^{(1)} / m_{kl}^{(1)}$ and $d_{kl}^{(2)} / m_{kl}^{(2)}$, to be similar. This can be assessed by contrasting the observed numbers of edges with their expectations conditional on the given margin totals, which is in line with the construction of Fisher's exact test on $2 \times 2$ contingency tables. In this regard, the theoretical random counterpart of $d_{kl}^{(1)}$ can be defined as
\begin{align*}
    D_{kl}^{(1)} &= \sum\limits_{\substack{i,j \\ j > i}} Y_{ij}^{(1)} \indFun{u_i^{(1)} \in [a_{k-1}, a_k)} \indFun{u_j^{(1)} \in [a_{l-1}, a_l)},
\end{align*}
for which under $H_0$ it approximately holds that
\begin{equation}
    \begin{gathered}
        D_{kl}^{(1)} \mid d_{kl} \sim \mbox{Hyp} \left( m_{kl}, d_{kl}, m_{kl}^{(1)} \right) \quad \mbox{with} \quad E_{kl}^{(1)} := \Ev (D_{kl}^{(1)} \mid d_{kl} ) = m_{kl}^{(1)} \frac{d_{kl}}{m_{kl}} \\
        \mbox{and} \quad V_{kl}^{(1)} := \Var (D_{kl}^{(1)} \mid d_{kl} ) = m_{kl}^{(1)} \frac{d_{kl}}{m_{kl}} \frac{m_{kl} - d_{kl}}{m_{kl}} \frac{m_{kl} - m_{kl}^{(1)}}{m_{kl} -1}.
    \end{gathered}
    \label{eq:notesNb1}
\end{equation}
Based on these specifications, we define our test statistic as
\begin{equation}
    T = \sum\limits_{\substack{k,l \\ l \geq k}} \frac{\left( D_{kl}^{(1)} - E_{kl}^{(1)} \right)^2}{ V_{kl}^{(1)} } \quad \mbox{with realization} \quad t = \sum\limits_{\substack{k,l \\ l \geq k}} \frac{\left( d_{kl}^{(1)} - E_{kl}^{(1)} \right)^2}{ V_{kl}^{(1)} }.
    \label{eq:testStat}
\end{equation}
Note that we only include the quantities of the first network due to the symmetry of the hypergeometric distribution. Moreover, summands for which $V_{kl}^{(1)} = 0$---resulting from $m_{kl}^{(1)}$, $m_{kl} - m_{kl}^{(1)}$, $d_{kl}$, or $m_{kl} - d_{kl}$ being zero---carry no relevant information and thus can simply be omitted from the calculation. In contrast, if $m_{kl}^{(1)}$ is large, $m_{kl}$ and $d_{kl}$ are large compared to $m_{kl}^{(1)}$, and $d_{kl} / m_{kl}$ is not close to zero or one, then $D_{kl}^{(1)}$ is known to be approximately normally distributed. Given that, we can conclude that
\begin{equation}
    T \stackrel{\text{a}}{\sim} \chi^2_{K (K+1) / 2}
    \label{eq:asymp}
\end{equation}
since, in this scenario, the test statistic is essentially the sum of squared (conditionally) independent random variables that approximately follow a standard normal distribution. If the latter condition does not apply, and assumption~\eqref{eq:asymp} is not reasonable to hold, we still can simulate a sample of the theoretical distribution by drawing $D_{kl}^{(1)} \mid d_{kl}$ according to (\ref{eq:notesNb1}) and calculating $T$ as in (\ref{eq:testStat}). In both cases, we can easily derive a critical value $c_{1-\alpha}$ to be compared with the realization $t$ of the test statistic. To do so, we pick the corresponding $(1-\alpha)$\-/quantile of either the theoretical distribution $\chi^2_{K (K+1) / 2}$ or the simulated sample. Finally, we reject null hypothesis~(\ref{eq:h0}) at the significance level of $\alpha$ if $t > c_{1-\alpha}$. The choice of an appropriate $K$ applied for these calculations is discussed in Section~\ref{sec:choiceK} of the Appendix. Note that altogether the presented test procedure follows a conception similar to the one underlying the log-rank test for time-to-event data.

Apparently, when conducting the test procedure on real-world networks, we obtain the joint graphon and the corresponding alignment of the networks by applying the estimation procedure described in Section~\ref{sec:graphonEst}. In the end, this enables us to appropriately approximate test statistic~(\ref{eq:testStat}). In this context, it is important to consider the general behavior of the joint graphon estimation under the alternative, that is, if hypothesis~(\ref{eq:h0}) does not hold. We stress that the intuition of the estimation procedure is to align the two networks as well as possible with respect to some suitable joint graphon model. Consequently, the expectation of $T$ will be higher the more the true graphons $w^{(1)}(\cdot,\cdot)$ and $w^{(2)}(\cdot,\cdot)$ differ after ``optimal'' alignment. This clearly implies that the power of our test is higher for instances that deviate more strongly from the null hypothesis.

\section{Applications}
\label{sec:Appl}
In this section, we showcase the applicability of the joint graphon estimation routine and the subsequent testing procedure. To give a comprehensive insight, this comprises both simulated and real-world networks. For an optimal estimation result and to best approximate test statistic (\ref{eq:testStat}), we repeat the estimation and testing procedure several times. In a modeling-oriented context, we would then typically pick the outcome with the lowest corrected $\operatorname{AIC}$. However, since here the focus is on the statistical testing aspect, we choose the estimation result which leads to the highest $p$-value, assuming that this provides an optimal lower bound for the outcome under the (potentially existing) oracle network alignment.

\subsection{Simulation Studies}
\subsubsection{Exemplary Application to Synthetic Data}
\begin{figure}%
	\centering
	\begin{minipage}[t]{1\textwidth}
		\centering
		\begin{minipage}[t]{0.48\textwidth}
			\centering
			\includegraphics[width=.99\textwidth]{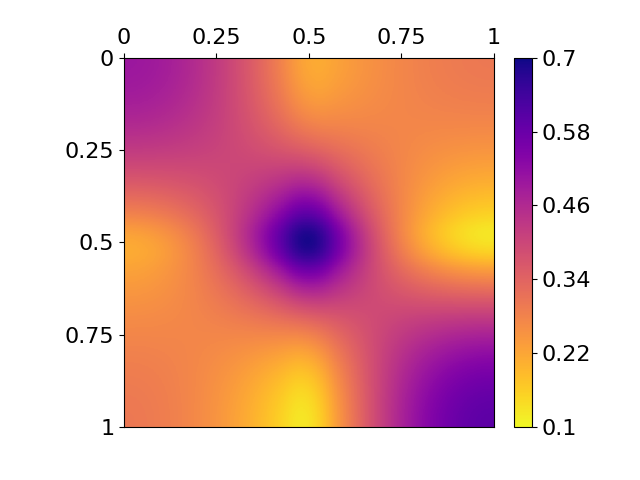}
		\end{minipage}
		\hfill
		\begin{minipage}[t]{0.48\textwidth}
			\centering
			\includegraphics[width=.99\textwidth]{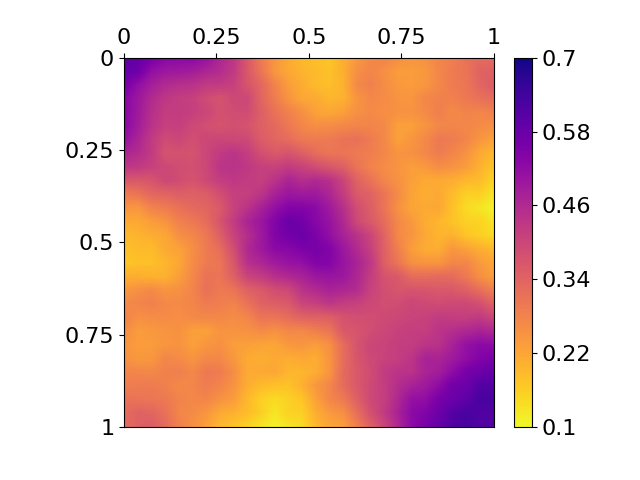}
		\end{minipage}
		\vfill
		
		\hspace*{-0.425cm}
		\begin{minipage}[c]{0.49\textwidth}
			\centering
			\includegraphics[width=.972\textwidth]{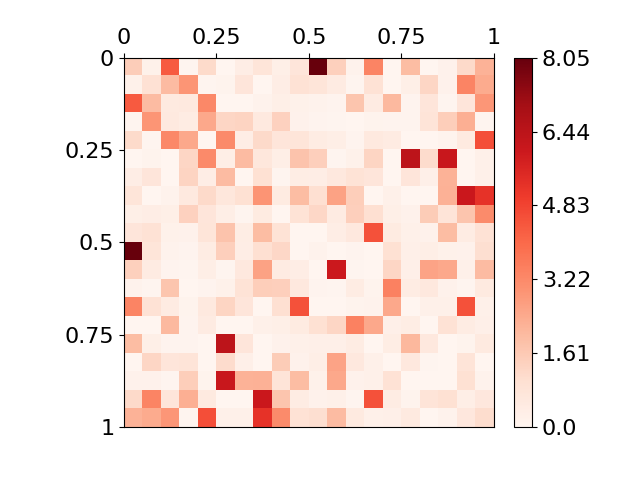}
		\end{minipage}
		\begin{minipage}[c]{0.49\textwidth}
		\hspace*{.096cm}
			\centering
			\includegraphics[width=.972\textwidth]{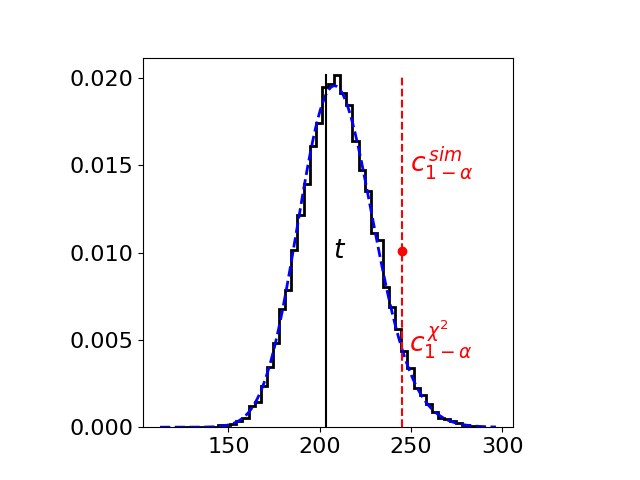}
		\end{minipage}
	\end{minipage}
	\caption{Joint graphon estimation for simulated networks with subsequent testing on equivalence of the underlying distribution models. The top row shows the true and the jointly estimated graphon on the left and right, respectively. The realizations of the terms of test statistic~(\ref{eq:testStat}), representing the dissimilarities of the two networks per rectangle, are visualized at the bottom left, where $m_{kl}^{(g)} \geq 100$ for $k \neq l$ and $\geq 45$ otherwise. The final result of the test statistic (black solid vertical line) as well as its distribution under $H_0$ are illustrated at the bottom right, where the black solid step function and the blue dashed curve depict the simulated and the asymptotic chi-squared distribution, respectively. The red dashed vertical lines visualize the critical values at a significance level of $5\%$, derived from the simulated (upper line) and the asymptotic distribution (lower line).
	}
	\label{fig:sim2GraEst1}
\end{figure}

To demonstrate the general capability of the joint graphon estimation and the performance of the subsequent testing procedure, we consider the graphon in the top left plot of Figure~\ref{fig:sim2GraEst1}. 
Its formation is inspired by and can be interpreted as a stochastic blockmodel with smooth transitions between communities. Based on this ground-truth model specification, we simulate two networks with $N^{(1)} = 200$ and $N^{(2)} = 300$ by making use of data-generating process~(\ref{eq:dataGen}). To recover the underlying structure, we then apply the presented EM-type algorithm, where, for initialization, we make use of an uninformative random node positioning. After several iterations, we achieve the reasonable joint graphon estimate at the top right, which fully captures the structure of the ground-truth graphon. Relying on the accompanying estimated node positions, we subsequently conduct the testing procedure on whether the underlying distributions are equivalent. To this end, we start with calculating the rectangle-wise differences according to the construction of test statistic~(\ref{eq:testStat}). The results are depicted as a heat map at the bottom left plot of Figure~\ref{fig:sim2GraEst1}. This reveals that the difference in the local edge density is rather low to moderate in most rectangles, whereas it is distinctly higher in a few others. Aggregating these differences yields a test statistic of $203.2$ as depicted by the black solid vertical line at the bottom right. Contrasting this result with the simulated $95\%$ quantile of the distribution of $T$ under $H_0$ as the critical value (red dashed vertical line) yields no rejection. Hence, the underlying distributions of the two networks do not significantly differ with respect to a significance level of $5\%$. As a final remark with regard to the bottom right plot, the simulated distribution of $T$ (black solid step function) and its theoretical approximation (blue dashed curve)---both relying on the assumption of $H_0$ being true---are very close to one another. Consequently, they also provide very similar critical values, namely $243.6$ and $244.8$, respectively. This demonstrates that asymptotic distribution~(\ref{eq:asymp}) represents a good approximation.

\subsubsection{Performance Analysis under $H_0$}
\begin{figure}%
	\centering
	\includegraphics[width=.95\textwidth]{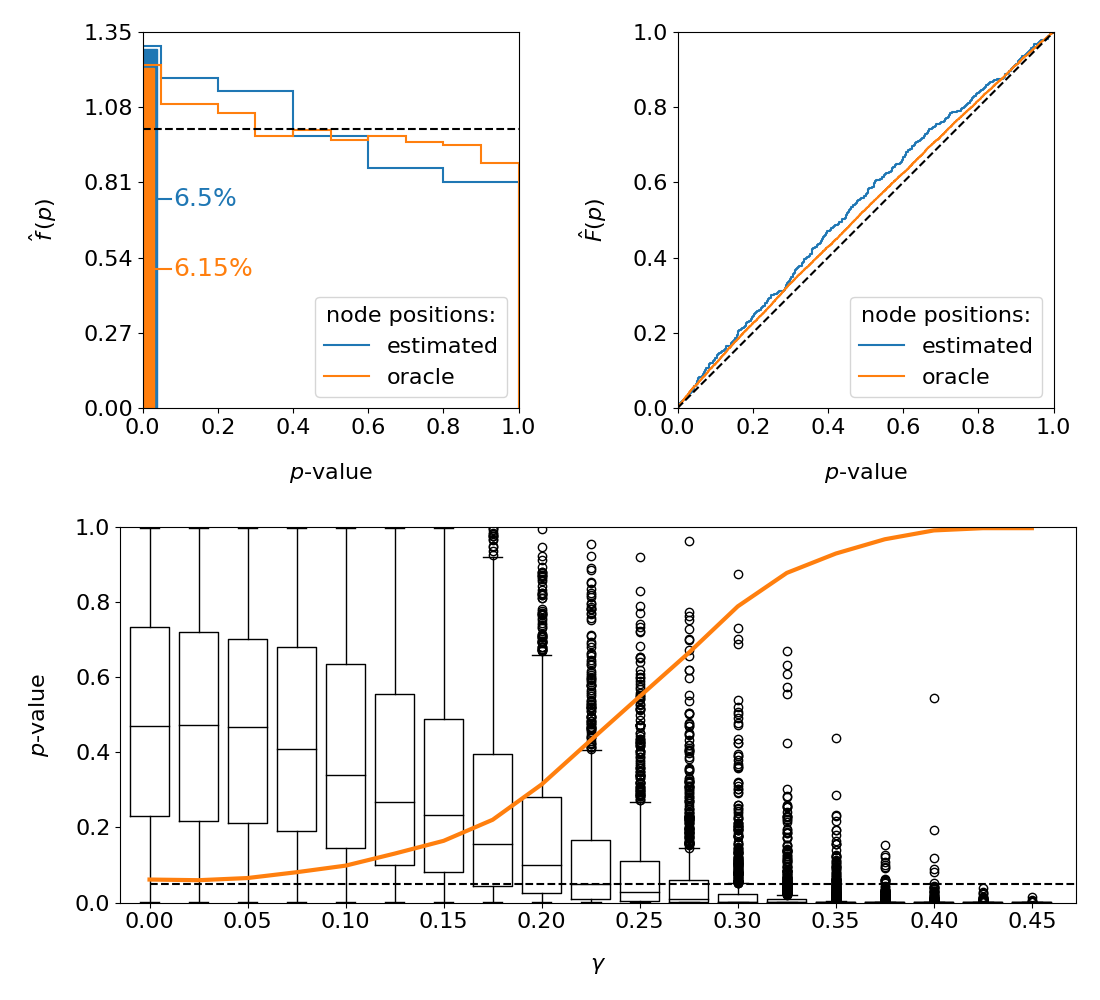}
	\caption{Performance of the testing procedure with regard to the resulting $p$-value; results are simulation-based. Top: empirical distribution of the $p$-value under $H_0$, illustrated as density and cumulative distribution function on the left and right, respectively. The black dashed lines illustrate the desired distributional behavior of an optimal test. Number of repetitions for estimated\, /\, oracle node positions: $400$\, /\, $10,000$. 
    Bottom: distribution of the $p$-value under $H_1$ and the usage of oracle node positions (in box plot format); based on $1,000$ repetitions each. The x-axis illustrates different settings according to formulation~(\ref{eq:makeH1}) (higher value of $\gamma$ implies stronger deviation from $H_0$). The black dashed horizontal line represents the $5\%$ significance level, and the orange curve illustrates the corresponding power.}
	\label{fig:compPVal}
\end{figure}

To evaluate the performance of the testing procedure in this example more profoundly, we repeat the above proceeding $400$ times, with newly simulated networks in each trial (remaining with $N^{(1)} = 200$ and $N^{(2)} = 300$). Note that we run the estimation procedure always ten times (with different random node positions as varying initialization) and finally pick the highest $p$-value as the actual result for the given network pair. These repetitions already provide a broad insight into the method's performance under the given setting. An even more extensive evaluation becomes possible when, in contrast to the proceeding above, the testing procedure is performed on the basis of the oracle node positions. This allows us to dramatically reduce the computational burden since it releases us from the preceding (computationally expensive) model estimation. As a consequence, we are able to increase the number of conducted tests to $10,000$. From these two repetition studies (using either $\hat{\bu}^{(g)}$ or $\bu^{(g)}$), we obtain rejection rates of $6.5\%$ and $6.15\%$ under the estimated and oracle node positioning, respectively. That means the test is slightly overconfident relative to the nominal significance level of $5\%$. The top row of Figure~\ref{fig:compPVal} 
shows additionally the empirical distributions of the observed $p$-values, illustrated as densities (left) and cumulative distribution functions (right). In accordance with the mildly inflated rejection rates, this exhibits a slight tendency to underestimate the $p$-value, i.e.\ interpreting the discrepancy as too high in distributional terms.

\subsubsection{Performance Analysis under $H_1$}
Conclusively, we are interested in evaluating the test performance under a false null hypothesis, which apparently requires formulating a suitable alternative. To this end, we ``shrink'' the heterogeneity within the graphon such that the present structure becomes less pronounced. The resulting graphon specification consequently tends more towards an Erd\H{o}s–R\'{e}nyi model, with the global density remaining unchanged. To be precise, based on $w^{(1)}(\cdot,\cdot)$, we formulate
\begin{align}
    w^{(2)}(u, v) := (1 - \gamma) \, w^{(1)}(u,v) + \gamma \, \bar{w}^{(1)}
    \label{eq:makeH1}
\end{align}
with $\gamma \in [0,1]$ and $\bar{w}^{(1)} = \iint w^{(1)}(u,v) \diff u \diff v$. Apparently, increasing the mixing parameter $\gamma$ leads to a stronger deviation from $H_0$. At the same time, this setting guarantees an optimal alignment of $w^{(1)}(\cdot,\cdot)$ and $w^{(2)}(\cdot,\cdot)$, meaning that there exists no rearrangement of $w^{(2)}(\cdot,\cdot)$ that is closer to $w^{(1)}(\cdot,\cdot)$ than specification~\eqref{eq:makeH1}. For this experiment, we again choose $N^{(1)} = 200$ and $N^{(2)} = 300$. Moreover, here we rely exclusively on the oracle node positions. This provides a lower bound of the power since the rejection rate can be expected to be higher when using estimated node positions instead (cf.\ previous analysis under $H_0$). The results for this setup are presented in the bottom plot of Figure~\ref{fig:compPVal}, where the distribution of the resulting $p$-value is illustrated for different settings of $\gamma$. The orange curve additionally visualizes the resulting power, i.e.\ the proportion of cases with $p < 0.05$. These results clearly show that the probability of detecting the false null hypothesis monotonically increases as the parameter $\gamma$ gets larger, which underpins the appropriateness of our test procedure.

Overall, the obtained simulation results demonstrate that the elaborated estimation and testing procedure yields reasonable results for assessing structural differences between networks. Building upon these findings, we next want to investigate the method's performance on real-world data.

\subsection{Real-World Examples}
\subsubsection{Facebook Ego Networks}
\begin{figure}%
	\centering
	\begin{minipage}[t]{1\textwidth}
		\centering
		\begin{minipage}[c]{0.49\textwidth}
			\centering
			\includegraphics[width=.99\textwidth]{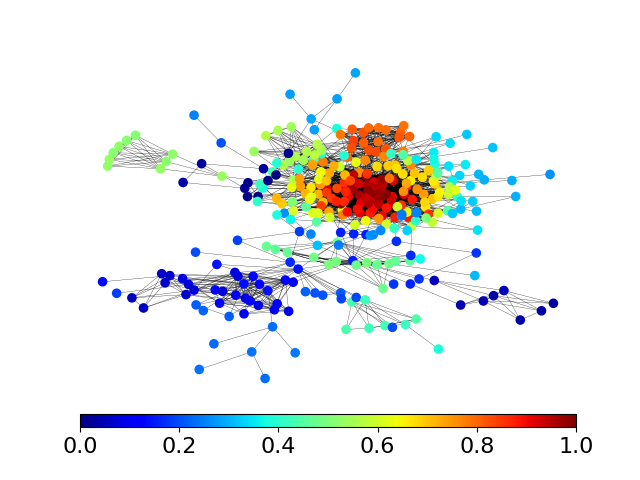}
		\end{minipage}
		\hfill
		\begin{minipage}[c]{0.49\textwidth}
			\centering
			\includegraphics[width=.99\textwidth]{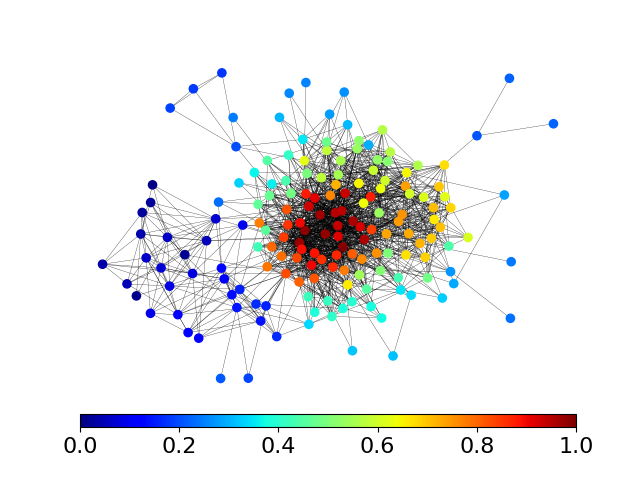}
		\end{minipage}
		\vfill
		\vspace*{0.2cm}

		\begin{minipage}[c]{0.49\textwidth}
			\centering
			\hspace*{-0.04cm}
			\includegraphics[width=.99\textwidth]{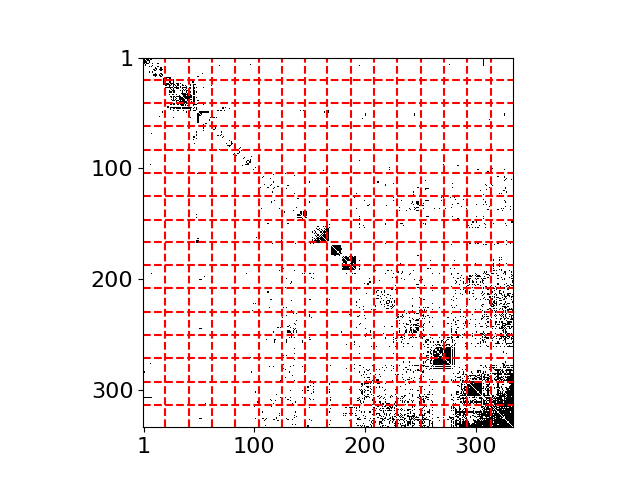}
		\end{minipage}
		\hfill
		\begin{minipage}[c]{0.49\textwidth}
			\centering
			\includegraphics[width=.99\textwidth]{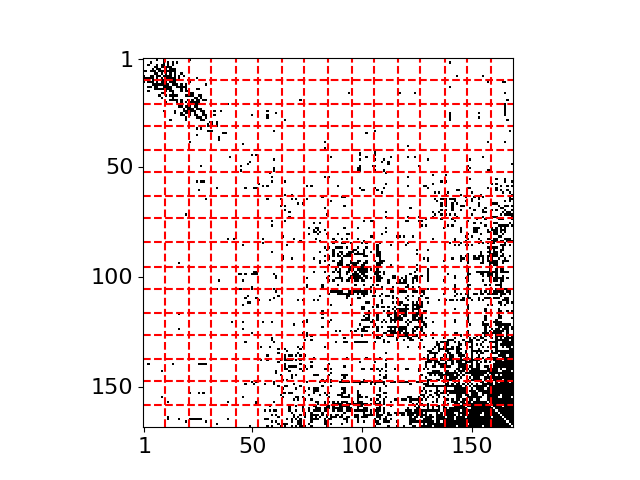}
		\end{minipage}
		\vfill
		\vspace*{0.4cm}
  
		\begin{minipage}[c]{0.49\textwidth}
			\centering
			\hspace*{0.175cm}
			\includegraphics[width=.99\textwidth]{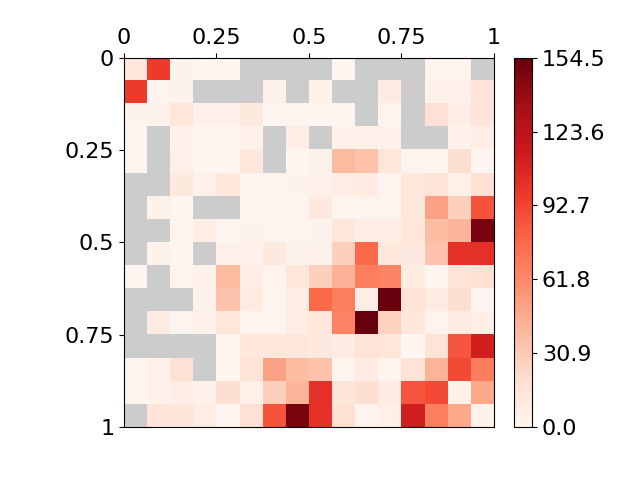}
		\end{minipage}
		\hfill
		\begin{minipage}[c]{0.49\textwidth}
			\centering
			\includegraphics[width=.99\textwidth]{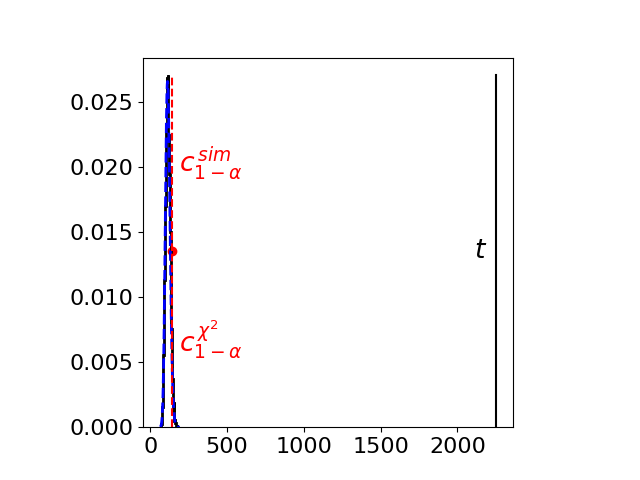}
		\end{minipage}
	\end{minipage}
	\caption{Comparison of two Facebook ego networks. Top: illustration of networks with coloring referring to estimated node positions. Middle: ordered adjacency matrices divided into blockwise segments. Bottom left: segment-wise differences between the two networks with $m_{kl}^{(g)} \geq 100$ for $k \neq l$ and $\geq 45$ otherwise; gray rectangles do not contain any observed edges ($d_{kl}=0$) and thus provide no information. Bottom right: realization of test statistic (black solid vertical line) plus corresponding distribution under $H_0$ (black solid step function and blue dashed curve represent simulated and asymptotic chi-squared distribution, respectively); critical values derived from the two types of distributions are represented by the upper and the lower red dashed vertical line.}
	\label{fig:realWorld1}
\end{figure}

As a first real-world example, we consider two Facebook ego networks which have been assembled by \cite{leskovec2012learning} and are publicly available on the Stanford Large Network Dataset Collection \citep{snapnets}. The two ego networks consist of $333$ and $168$ individuals, respectively, where the ego nodes are not included. An illustration of these networks is given in the top row of Figure~\ref{fig:realWorld1}, 
with nodes being colored according to resulting positions. In both networks, these final estimated node positions appear to be in line with the given network structure in terms of reflecting the nodes' embedding within the network. Moreover, they seem to be aligned across networks. For example, in both networks, the reddish nodes represent the rather centric individuals, whereas the nodes from the dark blue spectrum constitute a moderately interconnected branch that is more detached from the rest of the network. However, the segment-wise differences depicted at the bottom left in Figure~\ref{fig:realWorld1} exhibit some severe deviations. This can be clearly traced back to the blockwise division of the adjacency matrices as it results from partitioning the domain of edge positions (middle row). The aggregated differences ultimately result in a test statistic that is far from the sector of plausible values under the null hypothesis, as illustrated at the bottom right. Consequently, we can conclude that the structural behavior in the two networks differs significantly.

\subsubsection{Human Brain Functional Coactivation Networks}
\label{sec:humBrainCo}
In the second real-world application, we are concerned with differences in the human brain coactivation structure. To be precise, we compare two types of individuals, one with autism spectrum disorder (ASD) and the other with typical development (TD). In particular, we are interested in whether the functional connectivity within the brain significantly differs between these two groups (cf.\ the introductory example from Figure~\ref{fig:brainIntro}). For this analysis, we use resting-state functional magnetic resonance imaging data from the Autism Brain Imaging Data Exchange project \citep{abide}. More specifically, we employ preprocessed data provided by the Preprocessed Connectomes Project platform \citep{pcp}. Based on these person-specific datasets, we first calculate correlations between brain regions with respect to concurrent activation over time. Aggregating the results of participants from the same clinical group and employing an appropriate threshold finally yields the network-structured data which we aim to compare. To be precise, by performing the described preprocessing, we achieve for both groups, ASD and TD, a global connectivity pattern between $116$ prespecified relevant brain regions. Note that these regions are the same for both groups, which is why this could also be viewed as a comparison task under known node correspondence. However, we emphasize that in neurosciences, the transfer of competencies between brain regions is a well-known phenomenon, wherefore the general functional connectivity structure might be of greater relevance than the functional connection between specific regions. Further details on the acquisition and adequate transformation of the data are provided in Section~\ref{sec:detailBrain} of the Appendix.

\begin{figure}%
	\centering
	\begin{minipage}[t]{1\textwidth}
		\centering
		\begin{minipage}[c]{0.49\textwidth}
			\centering
			\includegraphics[width=.99\textwidth]{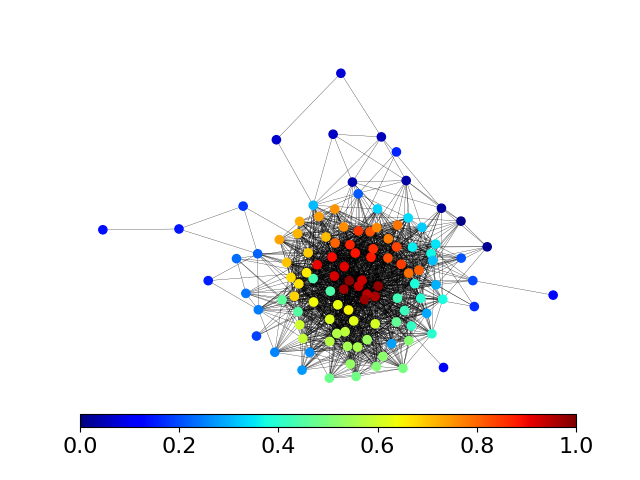}
		\end{minipage}
		\hfill
		\begin{minipage}[c]{0.49\textwidth}
			\centering
			\includegraphics[width=.99\textwidth]{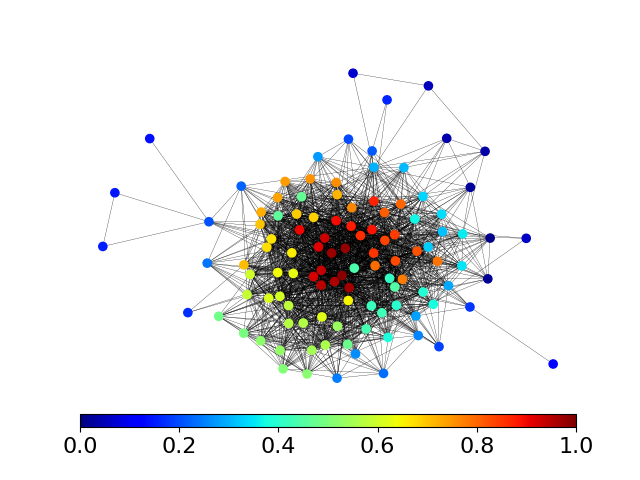}
		\end{minipage}
		\vfill
		\vspace*{0.2cm}
		
		\begin{minipage}[c]{0.49\textwidth}
			\centering
			\hspace*{-0.04cm}
			\includegraphics[width=.99\textwidth]{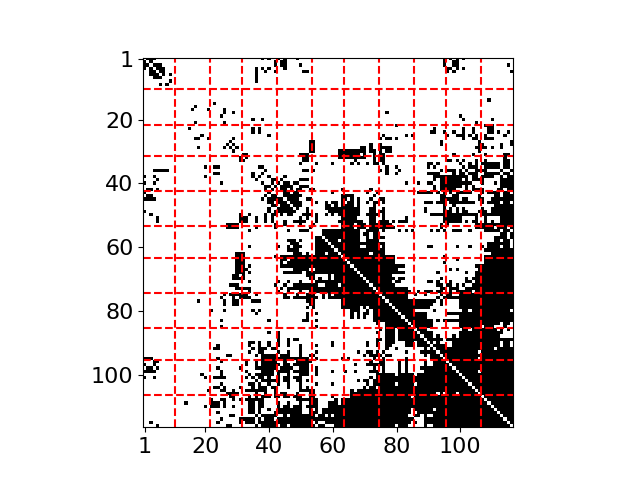}
		\end{minipage}
		\hfill
		\begin{minipage}[c]{0.49\textwidth}
			\centering
			\includegraphics[width=.99\textwidth]{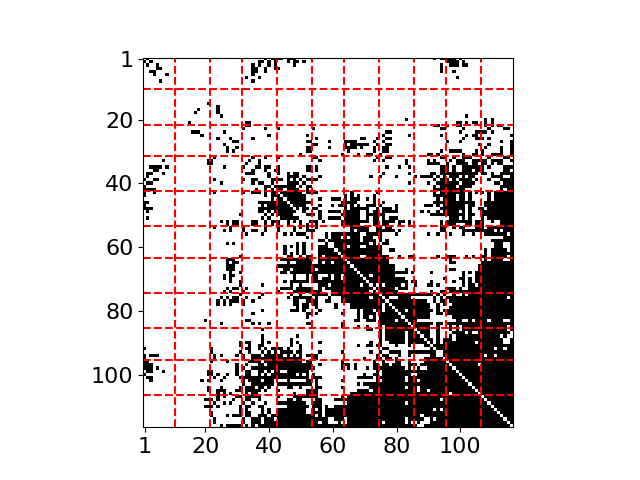}
		\end{minipage}
		\vfill
		\vspace*{0.4cm}
		
		\begin{minipage}[c]{0.49\textwidth}
			\centering
			\hspace*{0.175cm}
			\includegraphics[width=.99\textwidth]{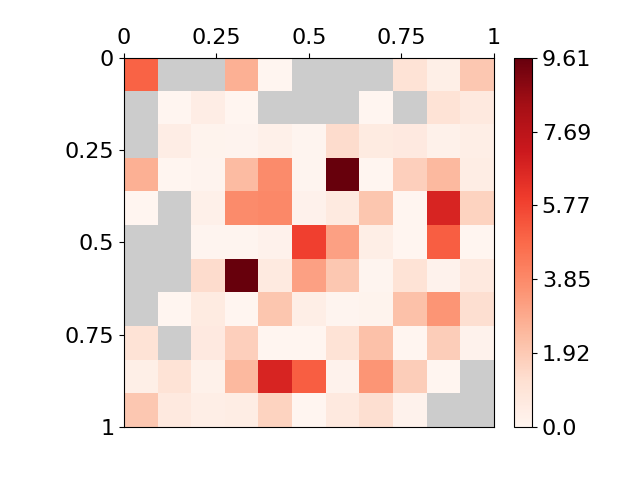}
		\end{minipage}
		\hfill
		\begin{minipage}[c]{0.49\textwidth}
			\centering
			\includegraphics[width=.99\textwidth]{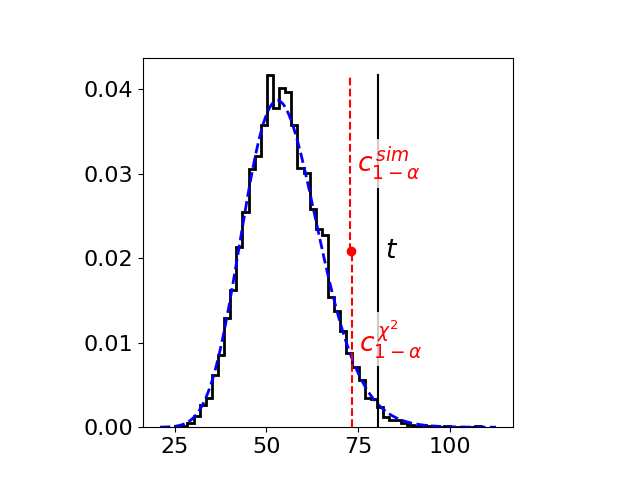}
		\end{minipage}
	\end{minipage}
	\caption{Comparison of functional coactivation in the human brain between groups of subjects with autism spectrum disorder and with typical development. The top row shows the networks of the ASD and the TD group on the left- and right-hand side, respectively. All illustration aspects are equivalent to the representation in Figure~\ref{fig:realWorld1}. The number of nodes per rectangle is again given by $m_{kl}^{(g)} \geq 100$ for $k \neq l$ and $\geq 45$ otherwise, where N/A's in the blockwise differences result from $d_{kl}$ or $m_{kl} - d_{kl}$ being zero.}
	\label{fig:realWorld2}
\end{figure}

For analyzing the differences in the brain coactivation structure between the two diagnostic groups, we again start with appropriately aligning the two networks. This is apparently done by employing the joint graphon estimation routine. The resulting node positions in relation to the embedding of nodes within the networks are illustrated in the top row of Figure~\ref{fig:realWorld2}. 
Again this reveals a plausible allocation of the nodes in the joint graphon model. The structural evolvement can further be evaluated by consulting the correspondingly ordered adjacency matrices (see middle row), where the red dashed lines represent the blockwise division resulting from the assignment of the edge positions to the rectangles in $[0,1]$. At first view, the formed structure looks quite similar in both matrices. Yet, on closer inspection, some blocks can be found where the density seems considerably different. This is also observed in the rectangle-wise differences depicted in the bottom left plot. In the end, aggregating these differences leads to rejecting the null hypothesis, as represented at the bottom right. To be precise, this test decision is based on a resulting $p$-value of $0.013$ (with reference to the simulated distribution under $H_0$).

Given that this outcome does not support an utterly unambiguous decision of the conducted test procedure, one might additionally be interested in the nature of the inferred differences. To address this, in Section~\ref{sec:brainLocDiff} of the Supplementary Material, we localize different behavior between the networks on the microscopic scale. Note that this can also be derived more or less directly from the joint graphon model. Moreover, for comparison reasons, we repeated the above analysis for two randomly selected disjoint subgroups of the TD group. According to the results illustrated in Figure~\ref{fig:realWorld2c} of the Supplementary Material, in this scenario, we do not observe a significant overall deviation. This further underlines the findings about the dissimilarity between the ASD and the TD group.

\section{Discussion and Conclusion}
\label{sec:discussion}
In the network comparison literature, the task of drawing statistical inference appears to be an open challenge up to now. We addressed this shortcoming in this paper by developing a nonparametric test on the equivalence of network structures. To do so, we utilized the smooth graphon model as a powerful tool for both describing and modeling the structure in complex networks. More precisely, extending previous estimation approaches towards a joint modeling framework allowed us to formulate a more generalized network alignment. Given that, local structure comparison can be carried out to uncover differences on the microscopic scale. Adequately aggregating these local differences finally enables to construct an appropriate nonparametric testing procedure on network data. Applying this comparison strategy to simulated and real-world networks clearly demonstrated its general applicability.

As outlined before, a crucial point for the proposed approach to work is the graphon model's property of decoupling structure and size. In the same line, one could think of further decoupling the global density by following the approach of \cite{bickel2009nonparametric}, i.e.\ by introducing network-related quantities $\rho^{(g)}$ that serve as individual density coefficients. Specifically, this means modifying formulation~(\ref{eq:dataGen}) by employing $\rho^{(g)} w^{(g)}(\cdot, \cdot)$, where $\hat{\rho}^{(g)}= [N^{(g)} (N^{(g)} -1)]^{-1} \sum_{i,j} y_{ij}^{(g)}$ could serve as an estimate that is independent of the rest of the structure. Such a framework consequentially might lead to a more balanced comparison strategy.

Beyond the applications presented in the previous section, which all refer to the situation with two networks, the method could easily be extended to cases with multiple or even a single network. For example, in the one-sample setting, to test whether a given network follows a hypothetical distribution $\Pb (\bY = \cdot \, ; \, w(\cdot, \cdot) , N )$, we could first align the network with the theoretical graphon. That is, applying the E-step based on $w(\cdot, \cdot)$. Given this alignment, we could then turn distributional assumption~(\ref{eq:notesNb1}) into a binomial distribution with the rectangle-specific mean over $w(\cdot, \cdot)$ as success probability, and, based on that, calculate the test statistic as in (\ref{eq:testStat}).

Besides the testing aspect, our approach could further be used to uncover relevant differences between networks on the microscopic scale. To be precise, determining the nodes or edges (present or absent) that contribute most to a quantified structural discrepancy between networks is possibly interesting in many situations. This is further elaborated in Section~\ref{sec:diffMicr} of the Supplementary Material.

As a last application case, the joint graphon estimation could further be used to predict edges \textit{between} separated networks by considering the cross-sample probabilities $w^{\text{joint}}(u_{i_1}^{(1)} , u_{i_2}^{(2)})$. This might be of interest when two (or more) networks are assumed to be samples of a larger global network. To the best of our knowledge, this has not been pursued by any other approach so far and hence constitutes a novel perspective. As a particular hurdle in this framework, the sampling strategy that is supposed for the drawing of subnetworks needs to be taken into account in the estimation routine. As far as this is not the Simple Induced Subgraph Sampling, where one selects a simple random sample of nodes within which all edges are observed, further adaptations are required. Hence, this lies beyond the scope of the paper.

\bigskip
\begin{center}
{\large\bf SUPPLEMENTARY MATERIAL}
\end{center}

\begin{description}

\item[Supplementary Manuscript:] (i) Description for deriving differences between networks at the microscopic level. As exemplary application, the two brain networks from Section~\ref{sec:humBrainCo} are considered.\\
(ii) Replication of the test on functional coactivation networks for two subgroups of the typical-development group (confer Section~\ref{sec:humBrainCo}).

\item[\texttt{Python}-package for testing on structural equivalence:] \texttt{Python}-package containing the code to perform the comparison methods described in the paper. The package also contains the preprocessed data of the human brain functional coactivation networks (see Section~\ref{sec:humBrainCo}). (GNU zipped tar file)

\end{description}

\bibliographystyle{chicago}
\bibliography{main.bib}

\appendix

\section*{Appendix}
\addcontentsline{toc}{section}{Appendices}
\renewcommand{\thesubsection}{\Alph{subsection}}

\subsection{Implementation of the Gibbs Sampler}
\label{sec:GibbsApp}
In an iterative joint graphon estimation procedure, the joint posterior distribution of the node positions given $w^{\text{joint}}(\cdot,\cdot)$ can be simulated by constructing a Gibbs sampler. We stress that the node positions are independent between networks and thus the Gibbs sampling procedure can be conducted for each network separately. The MCMC framework is then build upon full conditional distribution~(\ref{eq:gibbs}) and can be formulated as follows. For the successive updating procedure, we consider $\bu^{(g), \, <t>} = (u_1^{(g), \, <t>}, \ldots, u_{N^{(g)}}^{(g), \, <t>})$ to be the current state of the Markov chain. In the $(t+1)$-th step, component $i$ is then updated according to (\ref{eq:gibbs}), where all other components remain unchanged, i.e.\ $u_j^{(g), \, <t+1>} := u_j^{(g), \, <t>}$ for $j \neq i$. To do so, we propose a new position $u_i^{(g), \, *}$ by drawing from a normal distribution under the application of a logit link. To be precise, we first calculate
\begin{align*}
    v_i^{(g), \, <t>} = \operatorname{logit} (u_i^{(g), \, <t>}) = \log \left( \frac{u_i^{(g), \, <t>}}{1 - u_i^{(g), \, <t>}} \right),
\end{align*}
then we add a normal term in the form of $v_i^{(g), \, *} = v_i^{(g), \, <t>} + \operatorname{Normal}(0, \sigma_v^2)$, and finally we accomplish the retransformation through
\begin{align*}
    u_i^{(g), \, *} = \operatorname{logit}^{-1} (v_i^{(g), \, *}) = \frac{\exp (v_i^{(g), \, *})}{1 + \exp (v_i^{(g), \, *})}.
\end{align*}
In this setting, the variance $\sigma_v^2$ should be chosen such that a balance between a wide-ranging exploration and a high acceptance rate is achieved. Given these formulations, the proposal density can be written as
\begin{align*}
    q(u_i^{(g), \, *} | u^{(g), \, <t>}_{i}) =& \; \frac{\partial u^{(g), \, *}_i}{\partial z^{(g), \, *}_i} \phi(z^{(g), \, *}_i|z^{(g), \, <t>}_{i}) \\
    \propto& \; \frac{1}{u^{(g), \, *}_i (1-u_i^{(g), \, *})} \\
    & \;\cdot \exp \left(-\frac{1}{2} \frac{(\text{logit } (u_i^{(g), \, *}) - \text{ logit }(u^{(g), \, <t>}_{i}))^2}{\sigma^2}\right),
\end{align*}
which leads to a proposal ratio of
\begin{align*}
    \frac{q(u^{(g), \, <t>}_{i} | u^{(g), \, *}_i)}{q(u_i^{(g), \, *}|u^{(g), \, <t>}_{i})} = \frac{u^{(g), \, *}_i(1-u^{(g), \, *}_i)}{u^{(g), \, <t>}_{i} (1- u^{(g), \, <t>}_{i})}.
\end{align*}
In combination with the likelihood ration, the acceptance probability of the proposal, i.e.\ the probability for setting $u_i^{(g), \, <t+1>} := u_i^{(g), \, *}$, can be calculated through
\begin{align*}
    \min \left\{ 1, \quad \prod_{j \neq i} 
    \left[ 
    \vphantom{\left(  \frac{1-w(u^{(g), \, *}_i, u^{(g), \, <t>}_{j})}{1-w(u^{(g), \, <t>}_{i}, u^{(g), \, <t>}_{j})}   \right)^{1-y_{ij}}}
    \right. \right. & \left. \left. \left(  \frac{w(u_i^{(g), \, *}, u^{(g), \, <t>}_{j})}{w(u^{(g), \, <t>}_{i}, u^{(g), \, <t>}_{j})}  \right)^{y_{ij}} \right. \right.
    \\
    \cdot & 
    \left. \left.
    \left(  \frac{1-w(u^{(g), \, *}_i, u^{(g), \, <t>}_{j})}{1-w(u^{(g), \, <t>}_{i}, u^{(g), \, <t>}_{j})}   \right)^{1-y_{ij}} \right] \frac{u^{(g), \, *}_i(1-u^{(g), \, *}_i)}{u^{(g), \, <t>}_{i} (1- u^{(g), \, <t>}_{i})} \right\}.
\end{align*}
In case the decision yields a rejection of the proposal, we set $u_i^{(g), \, <t+1>} := u_i^{(g), \, <t>}$. Applying this updating strategy, which comprises the proposal of a new position plus the decision about its acceptance, to all $i = 1,\ldots, N^{(g)}$ completes one global update. Finally, we achieve a proper Gibbs sampling routine through consecutively repeating this global updating scheme. After cutting the burn-in period and applying an appropriate thinning, this approach yields a sample of the desired joint posterior distribution of the node positions.

\subsection{Derivative and Penalization of the B-Spline Function}
\label{sec:splineApp}
As has been show in Section~\ref{sec:spline}, the log-likelihood of a B-spline function can be straightforwardly extended towards the situation with multiple datasets. Given the formulation from (\ref{eq:logLik}), the score function can be calculated as
\begin{align*}
    \bs ( \btheta) &= \left[ \frac{\partial \ell (\btheta)}{\partial \btheta} \right]^\top \\
    &= \sum\limits_g \sum\limits_{\substack{i,j \\ j \neq i}} [ \bB_{ij}^{(g)} ]^\top \left( \frac{y_{ij}^{(g)}}{w_{\btheta}^{\text{joint}}(\hat{u}_i^{(g)},\hat{u}_j^{(g)})} - \frac{1-y_{ij}^{(g)}}{1-w_{\btheta}^{\text{joint}}(\hat{u}_i^{(g)},\hat{u}_j^{(g)})} \right).
\end{align*}
This, in turn, leads to the Fisher information in the form of
\begin{align*}
    \bF (\btheta) &= - \Ev \left( \frac{\partial \bs ( \btheta)}{\partial \btheta} \right) \\
    &= \sum\limits_g \sum\limits_{\substack{i,j \\ j \neq i}} [ \bB_{ij}^{(g)} ]^\top \bB_{ij}^{(g)} \left[ w_{\btheta}^{\text{joint}} \left( \hat{u}_i^{(g)},\hat{u}_j^{(g)} \right) \cdot \left( 1 - w_{\btheta}^{\text{joint}} \left( \hat{u}_i^{(g)}, \hat{u}_j^{(g)} \right) \right) \right]^{-1}.
\end{align*}
These results can then be used to implement the Fisher scoring procedure, where in (\ref{eq:penFuncs}) we additionally add a penalization term to guarantee a smooth estimation result. For penalizing ``neighborhood'' elements of the parameter vector $\btheta = \left( \theta_{11},\ldots, \theta_{1L}, \theta_{21}, \ldots, \theta_{LL} \right)^\top$, the penalization matrix can be formulated through
\[
    \bP = \left( \bJ_L \otimes \bI_L \right)^{\top} \left( \bJ_L \otimes \bI_L \right) +  \left( \bI_L \otimes \bJ_L \right)^{\top} \left( \bI_L \otimes \bJ_L \right),
\]
where $\bI_L$ is the identity matrix of size $L$ and
\[
    \bJ_L = 
    \begin{pmatrix}
    1       & -1                            & \phantom{-}0          & \multicolumn{2}{c}{\phantom{-}\cdots}        & \phantom{-}0 \\
    0       & \phantom{-}1                  & -1                    & \multicolumn{2}{c}{\phantom{-}\cdots}        & \phantom{-}0 \\
    \vdots  & \multicolumn{3}{c}{\ddots}                            &                                   & \phantom{-}\vdots \\
    0       & \multicolumn{2}{c}{\phantom{-}\cdots}    & \phantom{-}0          & \phantom{-}1                      & -1 \\
    \end{pmatrix} 
    \in \mathbb{R}^{(L-1) \times L}.
\]

\subsection{Choosing the Number and Extent of Rectangles}
\label{sec:choiceK}
In order to appropriately test null hypothesis~(\ref{eq:h0}), in Section~\ref{sec:testing}, we have developed an approach that relies on the partition of the graphon's domain. According to formulation~(\ref{eq:testQuants}), that involves the number of rectangles, $K$, as well as their concrete specification in the form of $[a_{k-1}, a_k) \times [a_{l-1}, a_l)$. In this regard, we emphasize that two aspects need to be observed. On the one hand, the joint graphon should be approximately constant within rectangles, requiring $[a_{k-1}, a_k) \times [a_{l-1}, a_l)$ to be not too expansive. On the other hand, the amount of edge variables per network falling into these blocks should be high, which needs rather broad rectangles. Thus, a trade-off between these two opposed requirements should be reached. In general, we choose $K$ to grow more slowly than both network dimensions, e.g.\ scaling as $\min_g \sqrt{N^{(g)}}$. Note that choosing $K=1$ would imply to test whether the two networks possess the same global density under the assumption of a joint Erd\H{o}s–R\'{e}nyi model. Having determined a suitable value for $K$, we then simply specify the boundaries of the rectangles through $a_k = k/K$ for $k=0,\ldots,K$. In combination with the subsequent adjustment of the latent positions as described in Section~\ref{sec:e-step}, which leads to equidistance of the estimates $\hat{u}_i^{(g)}$, a general lower bound for the amount of contained nodes per interval, $N_k^{(g)}$, can be derived. To be precise, we can formulate
\begin{multline*}
    N_k^{(g)} = \left\vert \left\{i \in \{ 1,\ldots, N^{(g)}\}: \frac{i}{N^{(g)} +1} \in [a_{k-1}, a_k) \right\} \right\vert %
    \geq \left\lfloor \frac{1}{K} (N^{(g)} +1) \right\rfloor,
\end{multline*}
where $\lfloor x \rfloor$ returns the largest integer smaller than or equal to $x$. Given that, $K$ could also be chosen such that, per network, a prescribed minimum amount of edge variables per rectangle ($N_k^{(g)} N_l^{(g)}$ for $l > k$ and $N_k^{(g)} (N_k^{(g)}-1)/2$ for $l=k$) is guaranteed.

As a final remark, we emphasize that with regard to the rectangle-based test statistic, it seems natural to alternatively apply a histogram estimator (\citealp{Chan2014} or \citealp{WolfeOl:14}). However, the smooth graphon estimation adapted from \cite{sischka2022b} considers a global node ordering which refers not only to separated intervals but to the entire domain of $[0,1]$. This consequently facilitates the iterative estimation procedure and thus leads to a more plausible and faster converging node positioning.

\subsection{Acquiring and Processing of Brain Functional Activation Data}
\label{sec:detailBrain}
The data we use for analyzing differences in the functional brain activation are originally provided by the Autism Brain Imaging Data Exchange project (\citealp{abide}, \citealp{di2014autism}). However, we make use of preprocessed data that are directly accessible through the Preprocessed Connectomes Project platform (\citealp{pcp}, \citealp{craddock2013neuro}). To be precise, we here apply the Connectome Computation System pipeline \citep{xu2015connectome} and the reduction to the Automated Anatomical Labeling atlas \citep{tzourio2002automated}. For each participant, this yields a dataset that consists of activity measurements over time for $116$ prespecified brain regions (a.k.a.\ regions of interest). Based on these temporal activity measurements, we calculate Pearson's correlation coefficient between all pairs of brain regions which, per participant, leads to the corresponding functional connectivity matrix (\citealp{song2019characterizing}, \citealp{subbaraju2017identifying}). For this analysis, we rely on the data from New York University, comprising $73$ ASD patients and $98$ TD subjects. For aggregating these connectivity patterns per clinical group, we apply Fisher's transformation to the pairwise correlation coefficients, calculate their mean for all pairs of brain regions, and finally retransform these means \citep{pascual2018evaluating}. This yields for both diagnostic groups a $116 \times 116$ weighted connectivity matrix which we binarize by employing a threshold of $0.4$. Based on that, the two final networks we obtain both possess a global density of about $30\%$. With regard to the choice of the threshold, \cite{song2019characterizing} have found that this has only minor effects when comparing the networks.

\end{document}


\if1\blind
{
  \maketitle
} \fi

\spacingset{1.9} %

\section{Detecting Differences on the Microscopic Scale}
\subsection{Methodological Construction}
\label{sec:diffMicr}
\begin{figure}%
	\centering
    \vspace*{-0.4cm}
    
    \makebox[\textwidth][c]{\includegraphics[width=1.01\textwidth]{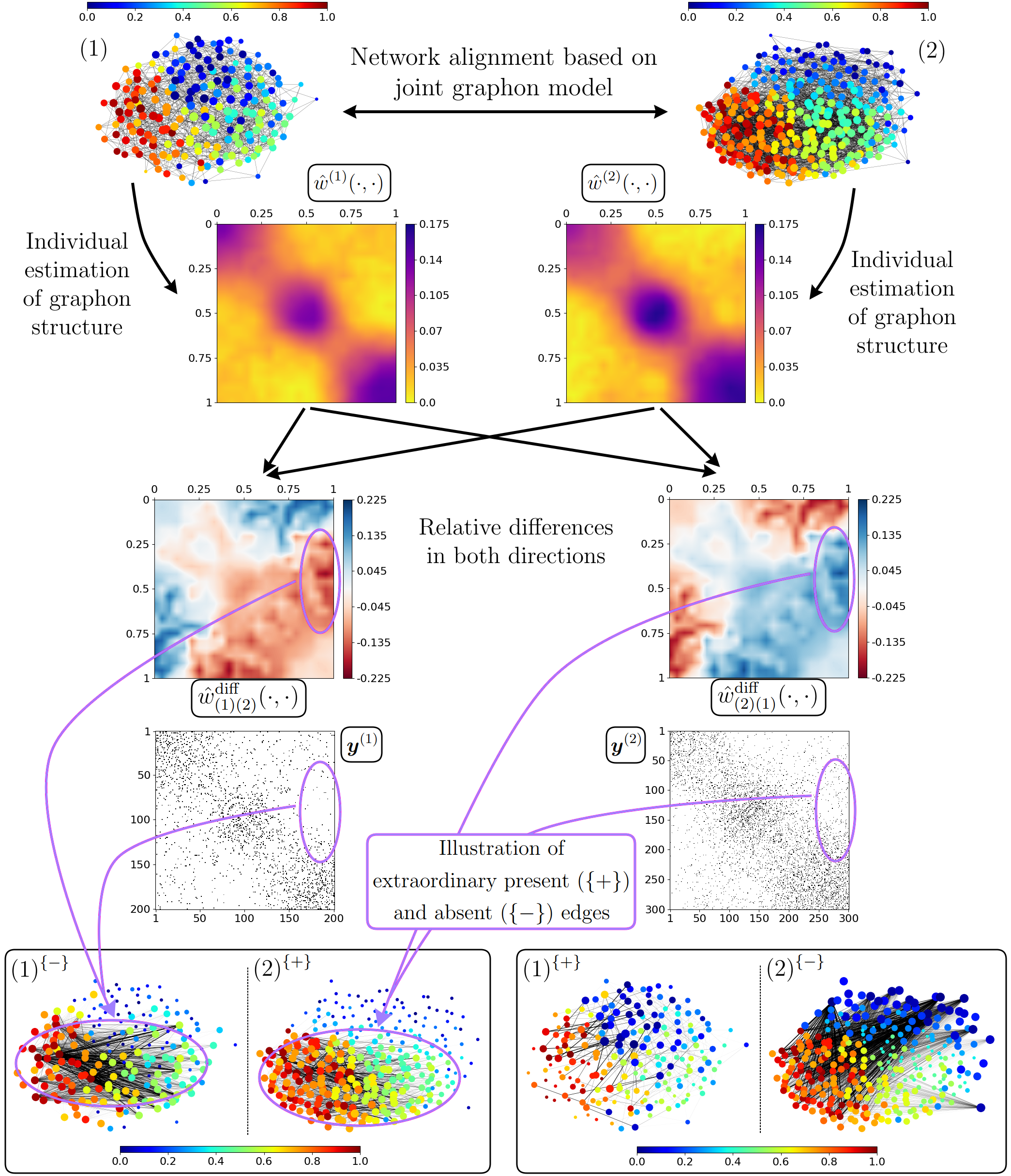}}
    \vspace*{-0.0cm}
    
    \hspace*{-1.15cm}
    \begin{minipage}{1.13\textwidth}
      \centering
      \caption{Detecting differences in networks at the microscopic level by employing the joint graphon-based alignment technique. The final network representations at the bottom row illustrate extraordinary \textit{absent} edges in network~1 and extraordinary \textit{present} edges in network~2 on the left and the converse on the right. The steps to get there are as follows: \mbox{(i) Align} networks based on joint smooth graphon estimation. (ii) Estimate individual graphons for disjoint networks separately. \mbox{(iii) Ca}l\-cu\-late relative differences between graphon estimates [$\rightarrow$ $\hat{w}_{(g_1)(g_2)}^{\text{diff}}(\cdot, \cdot)$]. (iv) Evaluating the function of relative differences at the edge positions provides information about contributions to the inferred structural deviation. This assessment is restricted to present or absent edges that are opposed to the formation of equivalent structures.
      }
	   \label{fig:micDiff}
    \end{minipage}
\end{figure}

In addition to the testing aspect, the network alignment based on the joint smooth graphon model also allows for determining structural differences on the edge level. In that regard, we are especially interested in differences that occur when inferring the structure separately. To address this, we first fit two separate graphon models to the two networks individually. To be precise, for this operation, we employ the node positions obtained from estimating the \textit{joint} graphon model and then perform the M-step as described in Section~\ref{sec:spline} of the paper but reduced to the use of only $\by^{(1)}$ or $\by^{(2)}$. This yields the separate estimates $\hat{w}^{(1)}(\cdot,\cdot)$ and $\hat{w}^{(2)}(\cdot,\cdot)$, respectively. Based on that, for $g_1,g_2 = 1, 2$ with $g_1 \neq g_2$, we calculate
\begin{align*}
    \hat{w}_{(g_1)(g_2)}^{\text{diff}}(u,v) = \frac{\hat{w}^{(g_1)}(u,v) - \hat{w}^{(g_2)}(u,v)}{\sqrt{\left\{ \hat{w}^{(1)}(u,v) \, [1 - \hat{w}^{(1)}(u,v)] + \hat{w}^{(2)}(u,v) \, [1 - \hat{w}^{(2)}(u,v)] \right\} / 2}}
\end{align*}
for $(u,v)^\top \in [0,1]^2$. With regard to data-generating process~(\ref{eq:dataGen}), this can be interpreted as the difference between expectations in relation to the averaged standard deviation. Hence, $\vert \hat{w}_{(g_1) (g_2)}^{\text{diff}}(\cdot, \cdot) \vert$ provides an appropriate measure to quantify the local differences between the found graphon structures. Moreover, it can be considered as a smoothed version of the impact on test statistic~(\ref{eq:testStat}). In turn, the contribution of the present or absent edge between node pair $(i, j)$ of network $g_1$ to the difference in the inferred structure can be quantified by evaluating
\begin{gather*}
    \begin{cases}
        \hat{w}_{(g_1) (g_2)}^{\text{diff},+}(\hat{u}_i^{(g_1)},\hat{u}_j^{(g_1)}) \, , & \text{if } y_{ij}^{(g_1)} = 1 \\
        \hat{w}_{(g_2) (g_1)}^{\text{diff},+}(\hat{u}_i^{(g_1)},\hat{u}_j^{(g_1)}) \, , & \text{otherwise}
    \end{cases} 
\end{gather*}
with $\hat{w}_{(g_1) (g_2)}^{\text{diff},+}(u,v) = \max \{ 0, \hat{w}_{(g_1)(g_2)}^{\text{diff}}(u,v) \}$. In particular, this means that the contribution is zero if the considered present or absent edge is contrary to the direction of the detected difference. Note that also here, the estimated node positions $\hat{u}_i^{(g_1)}$ are the ones stemming from the network alignment, meaning the positions resulting from estimating the joint graphon model. We stress that this approach is different from determining the deviation of present or absent edges from their ``transferred'' expectation, i.e.\ from simply calculating $\vert y_{ij}^{(g_1)} - \hat{w}^{(g_2)}(\hat{u}_i^{(g_1)}, \allowbreak \hat{u}_j^{(g_1)}) \vert$. Instead, here we aim to highlight those connections that, in one way or another, \textit{collectively} have enough impact to actually affect the inferred structure. To provide an illustrative intuition, this approach for detecting deviating behavior on the microscopic scale is additionally illustrated in Figure~\ref{fig:micDiff}. 
This representation allows to graphically demonstrate the single steps and thus to make the procedure much clearer.

\subsection{Application to Human Brain Coactivation Example}
\label{sec:brainLocDiff}
\begin{figure}%
	\centering
	\begin{minipage}[t]{1\textwidth}
		\centering
		\begin{minipage}[c]{0.49\textwidth}
			\centering
			\includegraphics[width=.99\textwidth]{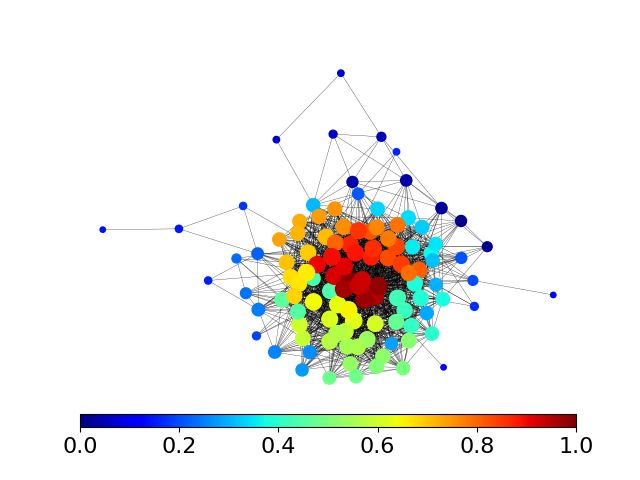}
		\end{minipage}
		\hfill
		\begin{minipage}[c]{0.49\textwidth}
			\centering
			\includegraphics[width=.99\textwidth]{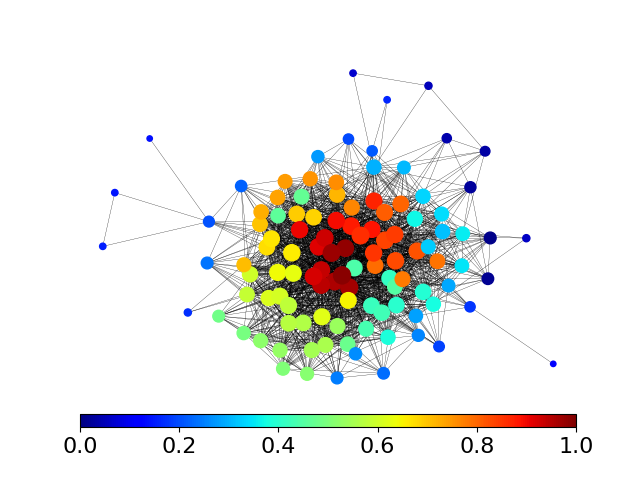}
		\end{minipage}
		\vfill
		\begin{minipage}[c]{0.49\textwidth}
			\centering
			\includegraphics[width=.99\textwidth]{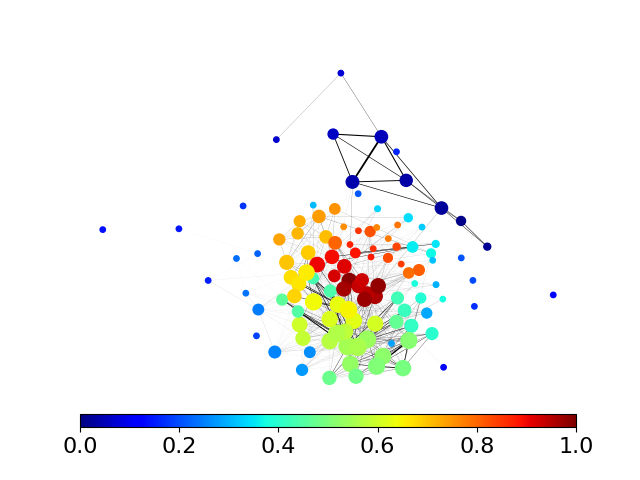}
		\end{minipage}
		\hfill
		\begin{minipage}[c]{0.49\textwidth}
			\centering
			\includegraphics[width=.99\textwidth]{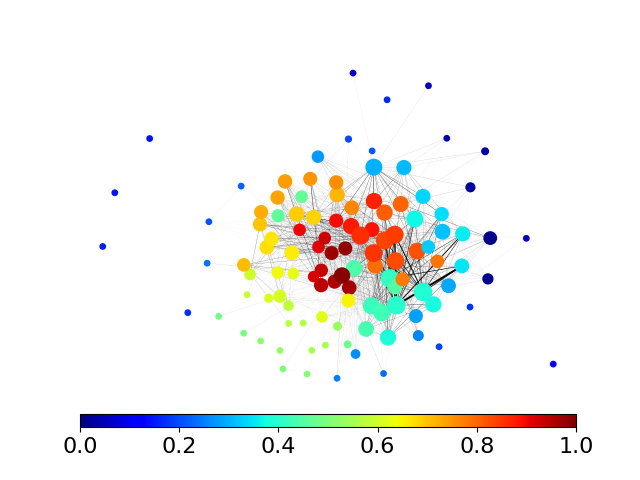}
		\end{minipage}
		\vfill
		
		\begin{minipage}[c]{0.49\textwidth}
			\centering
			\includegraphics[width=.99\textwidth]{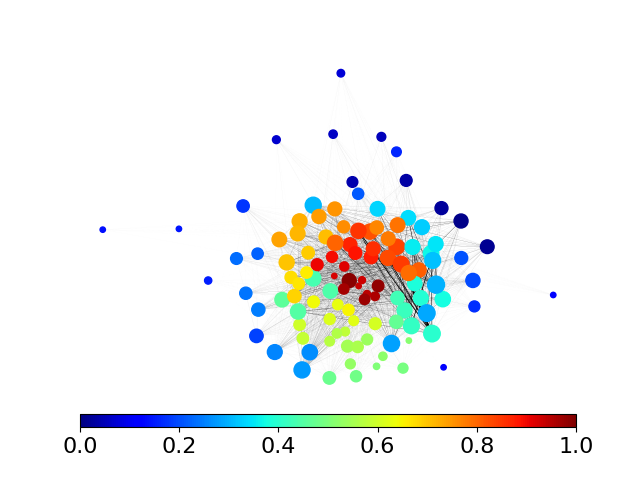}
		\end{minipage}
		\hfill
		\begin{minipage}[c]{0.49\textwidth}
			\centering
			\includegraphics[width=.99\textwidth]{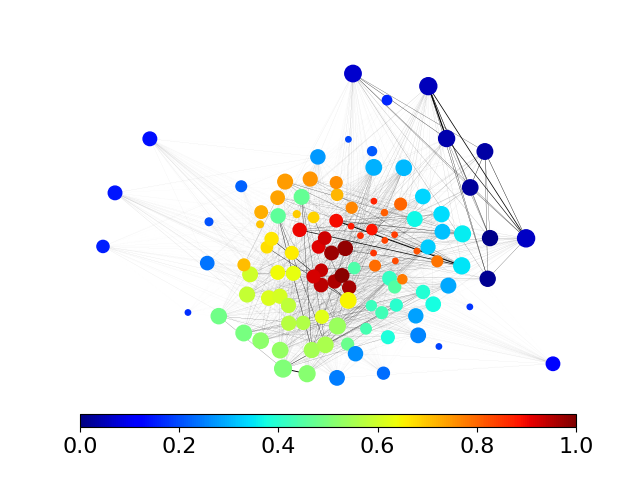}
		\end{minipage}
	\end{minipage}
	\caption{Localization of differences in the functional coactivation within the human brain. The results for the two clinical groups ASD and TD are represented in the left and the right column, respectively. Top: resulting connectivity between the $116$ considered brain regions after preprocessing; degree of nodes is illustrated by log-scaled node size. The lower four plots show observed present edges (middle) and absent edges (bottom) that form extraordinary structural patterns with respect to the structure found in the respective other network. The node sizes visualize the nodes' impact (log scale) as aggregation over the impact of attached edges.}
	\label{fig:realWorld2b}
\end{figure}

The analysis of the functional coactivation in the human brain has yielded a rather narrow test decision with regard to the two clinical groups, see Section~\ref{sec:humBrainCo} of the paper. Hence, we now additionally analyze the differences at the microscopic level. To quantify these differences, we make use of the approach described above. The corresponding results for this method are illustrated in Figure~\ref{fig:realWorld2b}, where the ASD and the TD group are represented on the left and right, respectively. In all these illustrations, the node coloring refers to the positions inferred through the joint graphon estimation procedure. For comparison reasons, the first row depicts again the functional connectivity networks just as obtained after preprocessing (cf.\ Figure~\ref{fig:realWorld2} of the paper). 
The second row shows the present edges in both networks which form collectives that are exceptional with respect to the structure uncovered from the respective other network. Here, the intensity of the depicted connections represents the magnitude of distinctiveness on an inverse log scale. The most extraordinary absent edges---also with respect to the structure of the respective other network and with magnitude represented by the inversely log-scaled intensity---are visualized at the bottom row. Based on these illustrations, we can conclude that, for example, the interconnection between the dark bluish nodes is much denser in the ASD group than in the TD group. Similar results are revealed for the interconnection between nodes from the green to the yellow color spectrum. In contrast, the connections between the dark orange and the cyan node bundle seem to be much more for the TD group. With regard to the test procedure carried out in the paper, these microscopic differences can be seen as a rough division of the calculated overall discrepancy represented in the form of the test statistic $t$.

Taking these analytical results together with the finding from the paper, we can (i) infer that the functional connectivity significantly differs between the ASD and the TD group and (ii) provide information on what these differences are composed of.

\section{Comparison of Brain Coactivation Between Typical-Development Subgroups}
\label{sec:showTDres}
\begin{figure}%
	\centering
	\begin{minipage}[t]{1\textwidth}
		\centering
		\begin{minipage}[c]{0.49\textwidth}
			\centering
			\includegraphics[width=.99\textwidth]{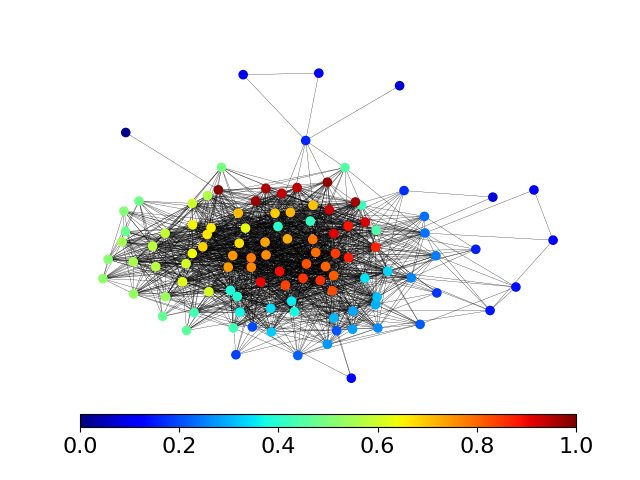}
		\end{minipage}
		\hfill
		\begin{minipage}[c]{0.49\textwidth}
			\centering
			\includegraphics[width=.99\textwidth]{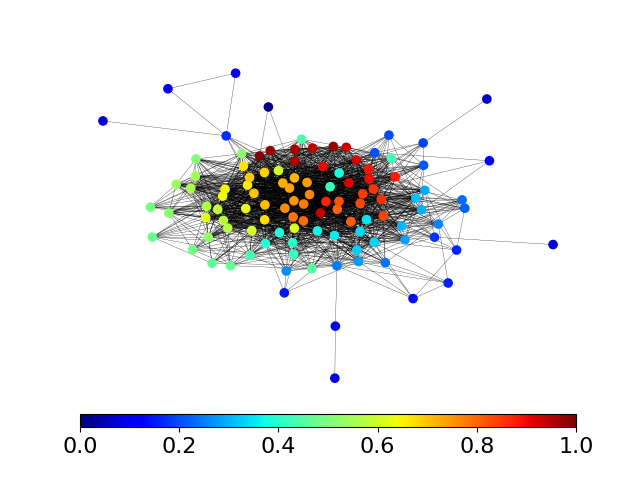}
		\end{minipage}
		\vfill
		\vspace*{0.5cm}
		
		\begin{minipage}[c]{0.49\textwidth}
			\centering
			\includegraphics[width=.99\textwidth]{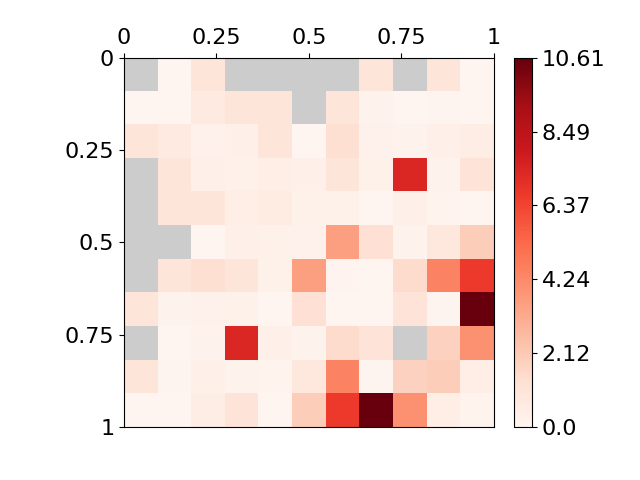}
		\end{minipage}
		\hfill
		\begin{minipage}[c]{0.49\textwidth}
			\centering
			\includegraphics[width=.99\textwidth]{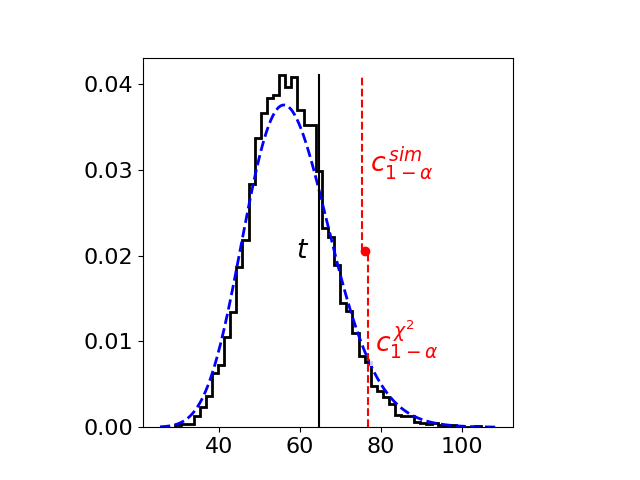}
		\end{minipage}
	\end{minipage}
	\caption{Comparison of the human brain functional coactivation between two subgroups of the TD group. Top: networks of subgroups with coloring referring to the inferred node positions. Bottom left: differences in rectangles according to the construction of test statistic~(\ref{eq:testStat}) in the paper. Bottom right: result of the test statistic, including the simulated and the theoretical distribution plus their corresponding critical values used for comparison.}
	\label{fig:realWorld2c}
\end{figure}

To further demonstrate that the found significant differences in the brain coactivation between the ASD and the TD group are meaningful (see Section~\ref{sec:humBrainCo} of the paper), we here repeat the analysis for two randomly selected subgroups of the TD group. The results are illustrated in Figure~\ref{fig:realWorld2c}. 
Considering the formation of the two networks in the top row, inclusive of the nodes' positional embedding found by the algorithm, this seems very similar for the two subgroups. This is also reflected by the differences in rectangles (bottom left) and the final realization of the test statistic (bottom right). Comparing the latter with the corresponding critical value shows that no significant difference on the global scale can be found.